\documentclass[12pt,a4paper]{article}

\usepackage[top=0.6in, bottom=1in, left=1in, right=1in, includeheadfoot]{geometry} 
\usepackage{graphicx}   
\usepackage{svg}       
\usepackage{fancyhdr}   
\usepackage{setspace}   
\usepackage[round, authoryear]{natbib}
\usepackage{url}
\usepackage[hidelinks,breaklinks]{hyperref}
\usepackage{float}
\usepackage{adjustbox}
\usepackage{xurl}
\usepackage{subcaption}
\usepackage{booktabs}
\usepackage{longtable}
\usepackage{amsmath}
\usepackage{dcolumn}
\usepackage{rotating}
\usepackage{enumitem}
\usepackage{dcolumn}     
\usepackage{soul}
\usepackage{url}

\newcolumntype{d}[1]{D{.}{.}{#1}}  
\setlist{itemsep=0.5pt, topsep=1pt}  

\fancypagestyle{plain}{%
   \fancyhf{}%
   \fancyhead[R]{\includegraphics[height=1.2cm]{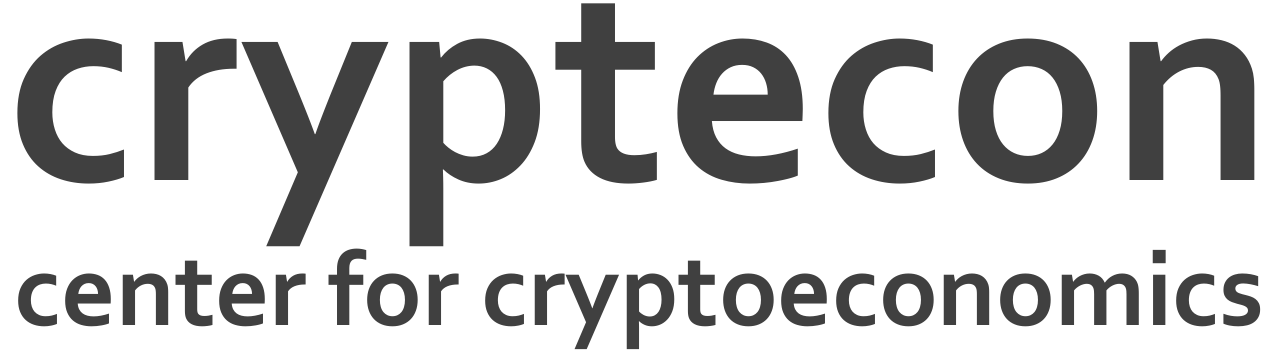}}%
   \fancyfoot[C]{\thepage}%
}

\title{Towards a Formal Framework of the Ethereum Staking Market\thanks{We thank Artem Kotelskiy for his valuable feedback and comments on this report. The research presented below is funded by the CyberFund MVI Grant program (\url{https://cyber.fund/content/mvi-grants}). The views expressed in the following are entirely our own.}}
\author{Juan Beccuti, Thunj Chantramonklasri, \\ Matthias Hafner, Nicolas Oderbolz \\ \\ \textit{Center for Cryptoeconomics}}   
\date{\today}   

\begin{document}

\maketitle              

\begin{abstract}
This paper examines how various categories of Ethereum stakers respond to changes in the consensus issuance schedule, and the potential impact of such changes on the composition of the staking market. To this end, we have develop and calibrate a game-theoretic model of the Ethereum staking market, incorporating strategic interactions between various staking agents. Our findings suggest that solo stakers may be more sensitive to variations in staking rewards than ETH holders using centralized exchanges or liquid staking providers. This increased sensitivity is driven not only by the cost structure of solo staking, but also by the competitive dynamics between different staking solutions in the market. When faced with a downward-sloping issuance schedule, staking agents compete for limited staking yields, and their choice of staking supply affects the revenues of other stakers. Therefore, the presence of other staking methods with access to MEV revenues and other DeFi yield sources when staking, as well as inattentive stakers, puts competitive pressure on solo stakers.  Consequently, our model predicts that a reduction in issuance is likely to crowd out solo stakers. We present preliminary empirical evidence to support this result, using an instrumental variable estimation approach to estimate the yield elasticity of staking supply for different staking categories.
\end{abstract}

\onehalfspacing         

\newpage
\pagenumbering{arabic}  
\setcounter{page}{1}    

\section{Introduction}\label{sec:introduction}

\subsection{Context}\label{sec:context}

The consensus issuance schedule, a central element of Ethereum's proof-of-stake (PoS) consensus protocol, determines how new tokens are created and distributed to validators as a reward for their participation in securing the network. To ensure robust network security, the issuance schedule must be structured to incentivize a sufficient number of validators to participate in staking. At the same time, if the issuance schedule sets rewards too high and facilitates levels of staking that exceed those required for economic security, the resulting inflation will unnecessarily dilute ETH holders. In addition, the resources devoted to running validator nodes could, on the margin, be allocated more efficiently elsewhere. To balance this trade-off, the issuance schedule follows a downward-sloping curve: when the total amount of staked ETH is low, rewards are high, but as more ETH is staked, rewards decrease. 

After the introduction of the Proof-of-Stake (PoS) consensus mechanism, the volume of staked ETH steadily increased and, after slightly declining in the past months, currently stands at approximately 35.8 million ETH or around 30\% of the total circulating supply.\footnote{Data obtained from \url{https://www.stakingrewards.com/asset/ethereum-2-0/analytics}, 16.07.2025} As a result of this growth, there has been speculation about potential long-term scenarios in which a significant portion of the circulating supply is locked within the staking protocol. In addition to an unbalanced trade-off between economic security and inflation, as well as an inefficient use of validator infrastructure, this could mean that in the long run, ETH is effectively replaced by more centralized and non-trustless liquid (re-)staking tokens as the predominant form of currency in the ecosystem. Motivated by these concerns, several proposals aimed at modifying the issuance schedule to effectively reduce staking incentives set by the protocol have emerged.\citep{drake2024practical, elowsson2024endgame, elowsson2024properties, elowsson2024tempered}

However, a change in the consensus issuance schedule could affect different types of stakers in different ways, which could have significant consequences for the composition of the Ethereum validator set, and therefore the degree of decentralization of the protocol. There exist different staking solutions in the Ethereum staking market, each with distinct implications for the protocol’s decentralization \citep{artem2024maximum}. While individual solo stakers contribute significantly to decentralization, centralized exchanges control substantial portions of staked ETH within proprietary infrastructures, introducing centralization risks. Liquid staking solutions, like Lido, tend to occupy an intermediate position: they typically feature decentralized governance structures and distribute their stake across multiple uncorrelated staking service providers. Liquid staking tokens benefit from strong network effects, which can enable a single issuer to gain significant market share and, consequently, substantial influence within the Ethereum ecosystem \citep{neuder2023magnitude}.

Some argue that reducing the staking rewards issued by the protocol would better balance the trade-off between economic security and inflation. It would also limit the continued adoption of liquid staking tokens (LSTs) and thus reduce the influence of liquid staking protocols. However, opponents maintain that Ethereum’s issuance policy should primarily serve to maximize the network’s economic security. They caution that reducing staking rewards could push solo stakers and smaller operators, who face higher setup costs and lack the economies of scale enjoyed by larger operators, out of the market. Consequently, staking could become more centralized if consensus issuance is reduced, which would ultimately weaken the protocol’s economic security and long-term stability \citep{artem2024maximum}.

\subsection{Research Questions and Methodology}\label{sec:research_questions}

The present research aims to contribute to the above discussion by developing a formal understanding of how different players in the Ethereum staking market form staking decisions and how these may be affected by changes in the issuance schedule. Specifically, two main research questions are addressed sequentially:
\begin{enumerate}
    \item \textit{How can we characterize the staking supply from different categories of stakers?}
    \item \textit{How does a change in the issuance schedule affect the staking supply of these stakers?}
\end{enumerate}   

\noindent To address these research questions, we review the relevant existing literature and provide a short overview of the staking market. We then develop a formal model describing the Ethereum staking market equilibrium in terms of the costs and revenues faced by individual agents within the ecosystem. By conducting numerical simulations of this model, we aim to understand the factors that drive staking behavior and consequently equilibrium staking outcomes. Additionally, we calibrate the model using specific parameter values that try to approximate the current market environment, evaluating how specific changes in the issuance schedule may affect staking market equilibrium.

\subsection{Summary of Main Findings}\label{sec:main_findings}

Using a combination of empirical evidence and a calibrated theoretical model, we demonstrate that solo stakers may be more susceptible to fluctuations in consensus issuance yields than ETH holders who stake through centralized exchanges or liquid staking providers. Our findings suggest that a reduced issuance schedule could lead to a reduced market share of solo staking and an increase in the market share of centralized exchanges, leading to greater centralization.

In our model, we begin by assuming that the cost structures of solo stakers alone would suggest lower supply elasticity in response to changes in the consensus issuance yield, compared to other staking methods. Even so, we find that other factors influence the staking decisions of solo stakers within the market, on net resulting in the staking supply of solo stakers being relatively sensitive to changes in the issuance schedule. Importantly, their staking supply decision is also shaped by market dynamics. Given a downward-sloping issuance schedule, stakers compete for issuance yields. A reduced issuance intensifies this competition further. Since other staking methods benefit from superior MEV access and DeFi yields, they are less affected by the change in issuance. This creates a crowding-out effect for solo stakers. Strategic solo stakers internalize these competitive pressures, making them more sensitive to shifts in consensus rewards.  

Our results emphasize the importance of policy measures such as MEV burn in mitigating the negative impact of issuance reductions on solo staking. Moreover, our model predicts that reduced issuance disproportionately reduces the profitability of solo staking compared to other staking categories. While beyond the scope of our model, this may lower long-term incentives for solo staking and may drive exits or shifts in supply towards liquid staking solutions, which can offer stakers additional DeFi yields.

\subsection{Structure}\label{sec:structure}

This paper is structured as follows: Section \ref{sec:lit_review} reviews the relevant existing literature. Section \ref{sec:empirical_evaluation} provides a brief data-driven overview of the Ethereum staking market. Section \ref{Section_Model} introduces a game-theoretic model of the Ethereum staking and outlines its main assumptions. Additionally, this section reports on the numerical simulation of the model, as well as the comparitive analysis under different issuance schedules. In Section \ref{empirical analysis}, we estimate the sensitivity of the staking supply of different staking categories to changes
in staking returns using an instrumental variable approach. Section \ref{discussion} discusses the results presented in the previous sections and Section \ref{conclusion} concludes.

\newpage

\section{Literature Review}\label{sec:lit_review}

First, \cite{Eloranta} conduct an empirical analysis of the Ethereum staking market in view of a potential change in the consensus issuance schedule. They identify a significant relationship between staking operation size and variations in staking returns, with historical data indicating that larger staking pools consistently achieve higher returns compared to single-validator pools. Moreover, they find that solo stakers demonstrate greater sensitivity to relative declines in yield compared to other staking categories. They suggest that this heightened responsiveness may indicate that individual validators would be disproportionately impacted by changes in network incentives.

\cite{ethIssuanceDiscovery} conduct interviews with various stakeholder groups in the Ethereum staking market. Based on these interviews, they argue that solo stakers exhibit heightened sensitivity to changes in the issuance curve relative to other stakeholders, a finding consistent with \cite{Eloranta}. Moreover, the authors find that large crypto holders generally regard staking rewards as a secondary benefit, with their primary focus being on ETH price appreciation. These individuals also tend to prefer liquid staking solutions for the utility of their LSTs provide in DeFi. Additionally, the study argues that retail investors tend to prioritize convenience and brand trust over short-term price fluctuations or minor yield variations. As a result, they often stake through platforms like Coinbase and are relatively insensitive to fluctuation in staking yields. Lastly, institutional stakeholders tend to emphasize security and regulatory compliance in their staking decisions, with their primary motivation being the preservation of their ETH holdings against dilution.

Overall, the available empirical literature recovers several key insights. First, solo stakers are consistently found to demonstrate greater sensitivity to fluctuations in staking yields compared to other staking categories. This finding challenges the conventional assumption that solo stakers are primarily motivated by altruism and a commitment to Ethereum's decentralization, rather than financial incentives. Furthermore, the literature suggests that the Ethereum staking market is to some degree segmented along different staking solutions. In particular, institutions and retail investors tend to concentrate around the staking services of centralized exchanges, such as Coinbase, while DeFi-native stakers tend to prefer liquid staking solutions like Lido.

Regarding the modeling approach adopted in this study, perhaps the most relevant existing work is that of \cite{julianmasolo}, who models the staking allocation process of ETH holders across different staking methods as a function of both monetary returns and non-monetary preferences, such as convenience, trust, and commitment to decentralization. Under the assumption of linear cost functions, the model predicts that changes in the issuance schedule do not affect an individual’s optimal choice of staking method. The result is that the overall market share of a given staking method remains unchanged when the issuance schedule is adjusted. While this modeling approach represents an important first step in modeling the distribution of staking supply in relation to issuance levels and the costs associated with different forms of staking, the results run counter to the empirical findings above. In particular, the observed higher sensitivity of solo stakers to changes in staking yields suggests that their market share would likely decline in response to reductions in issuance. Consequently, relaxing the assumption of linear cost functions may be important to achieving better alignment with the empirical evidence. Among other novel features, the theoretical model developed in the following takes a step in this direction by allowing for non-linear staking cost structures.

\section{Historical Staking Supply}\label{sec:empirical_evaluation}

\begin{figure}[H]
    \centering
    \caption{Staked ETH by staking method}
    \label{fig:ETH_Staked}
        \centering
        \includegraphics[width=0.8\linewidth]{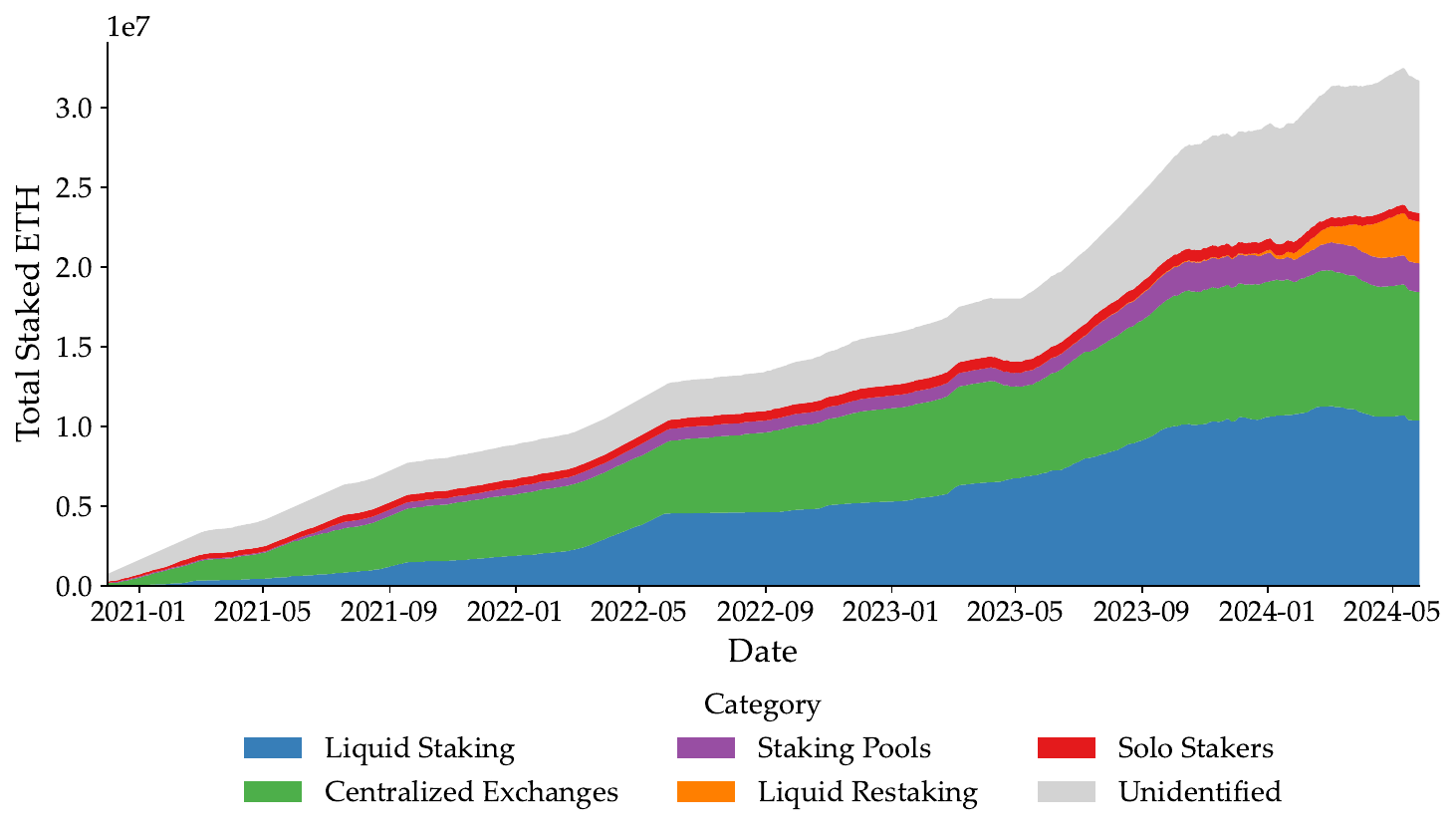}
    \caption*{\footnotesize \textit{Source}: Dune query by user hildobby, see \url{https://dune.com/queries/1941407/3202651}.}
\end{figure}

\begin{figure}[H]
    \centering
    \caption{Share of staked ETH by staking method}
    \label{fig:ETH_Staked_share}
        \centering
        \includegraphics[width=0.8\linewidth]{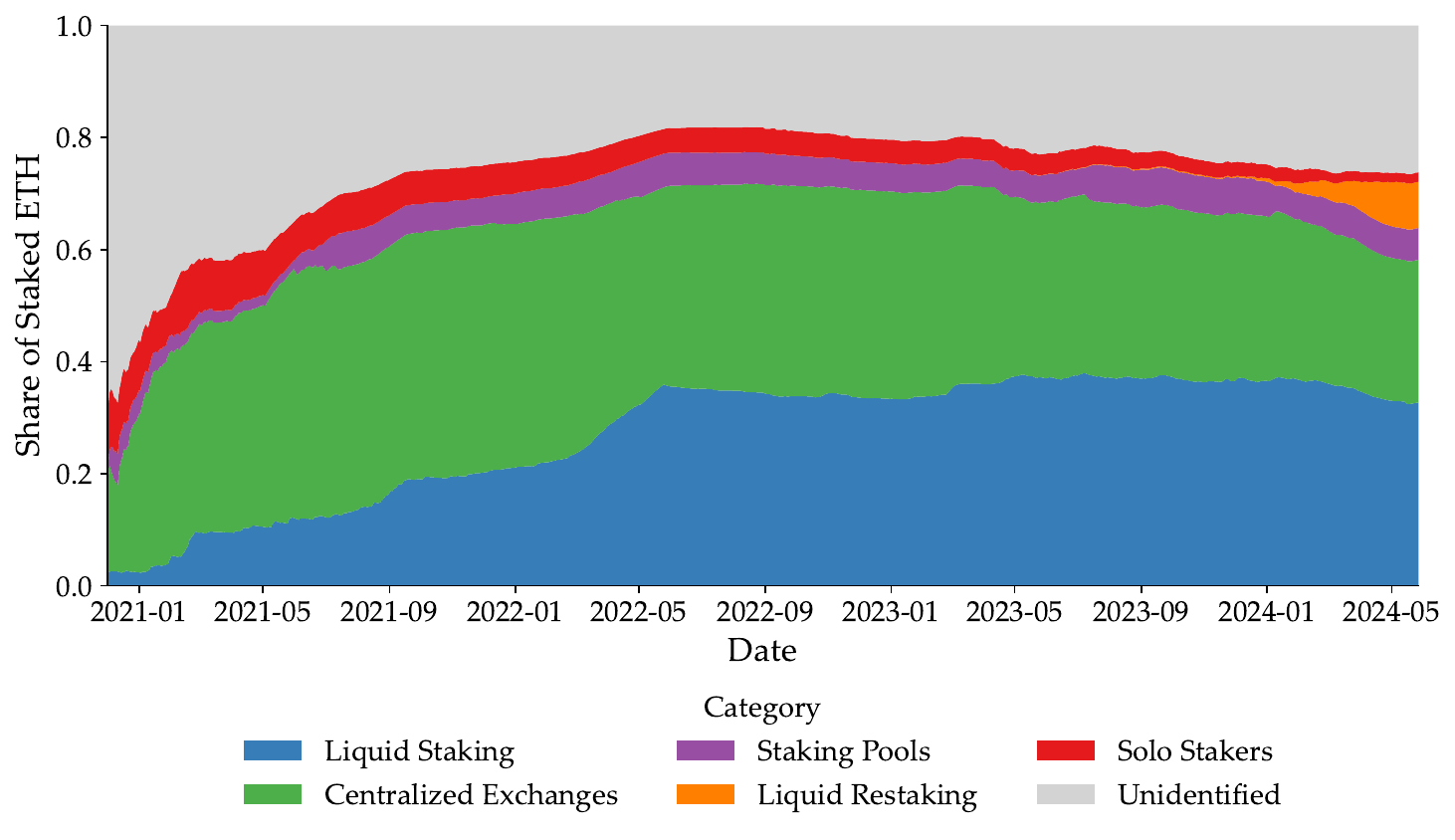}
    \caption*{\footnotesize \textit{Source}: Dune query by user hildobby, see \url{https://dune.com/queries/1941407/3202651}}
\end{figure}

\begin{figure}[H]
    \centering
    \caption{Validator entry and exit over time}
    \label{fig:EntryExit}
    \begin{subfigure}{0.88\linewidth}
        \centering
        \caption{Total}
        \includegraphics[width=\linewidth]{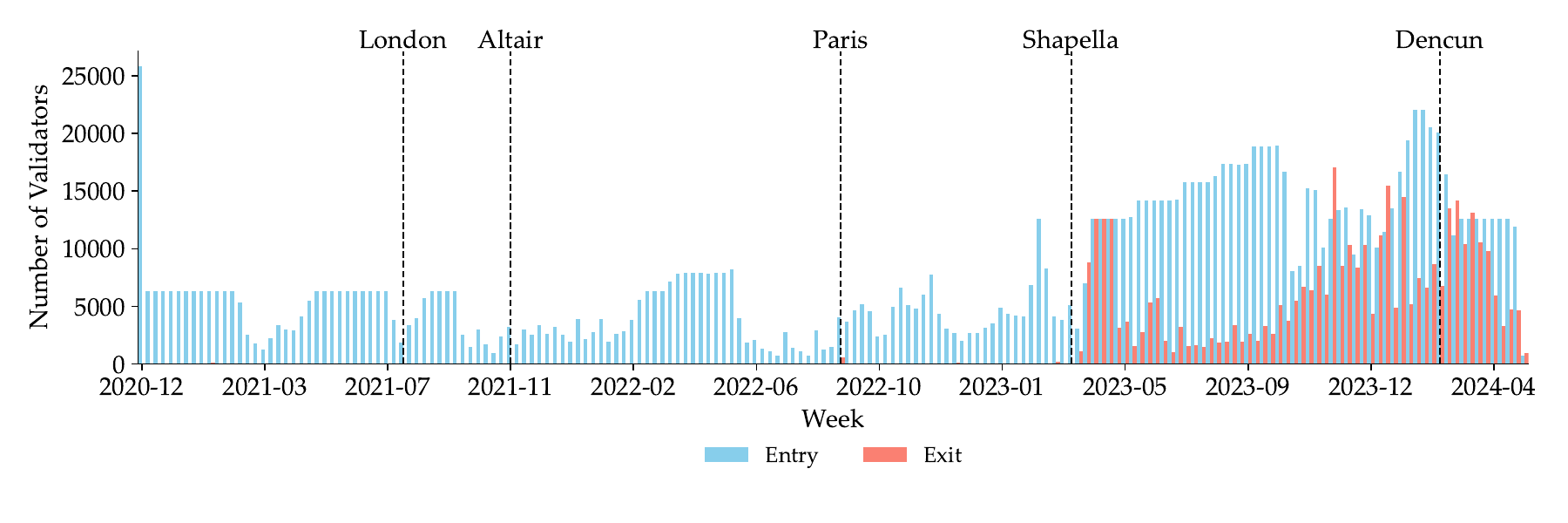}
    \end{subfigure}
    \hfill
        \begin{subfigure}{0.88\linewidth}
        \centering
        \caption{Solo Staking}
        \includegraphics[width=\linewidth]{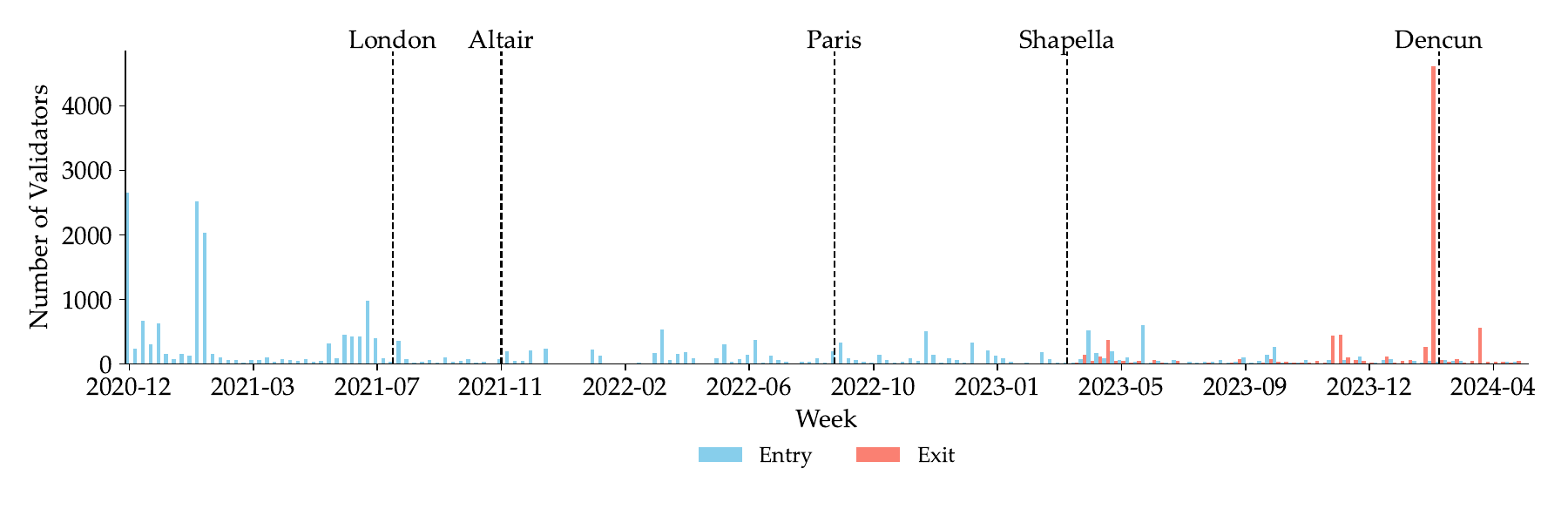}
    \end{subfigure}
    \hfill
        \begin{subfigure}{0.88\linewidth}
        \centering
        \caption{CEX}
        \includegraphics[width=\linewidth]{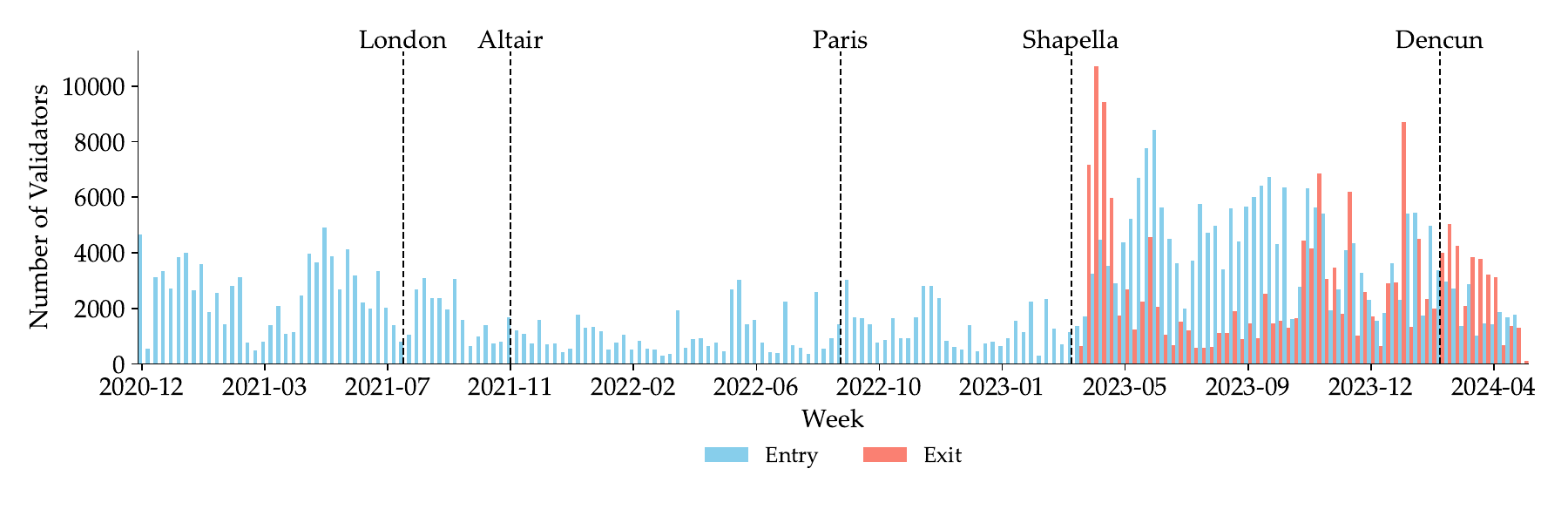}
    \end{subfigure}
    \hfill
        \begin{subfigure}{0.88\linewidth}
        \centering
        \caption{Liquid Staking}
        \includegraphics[width=\linewidth]{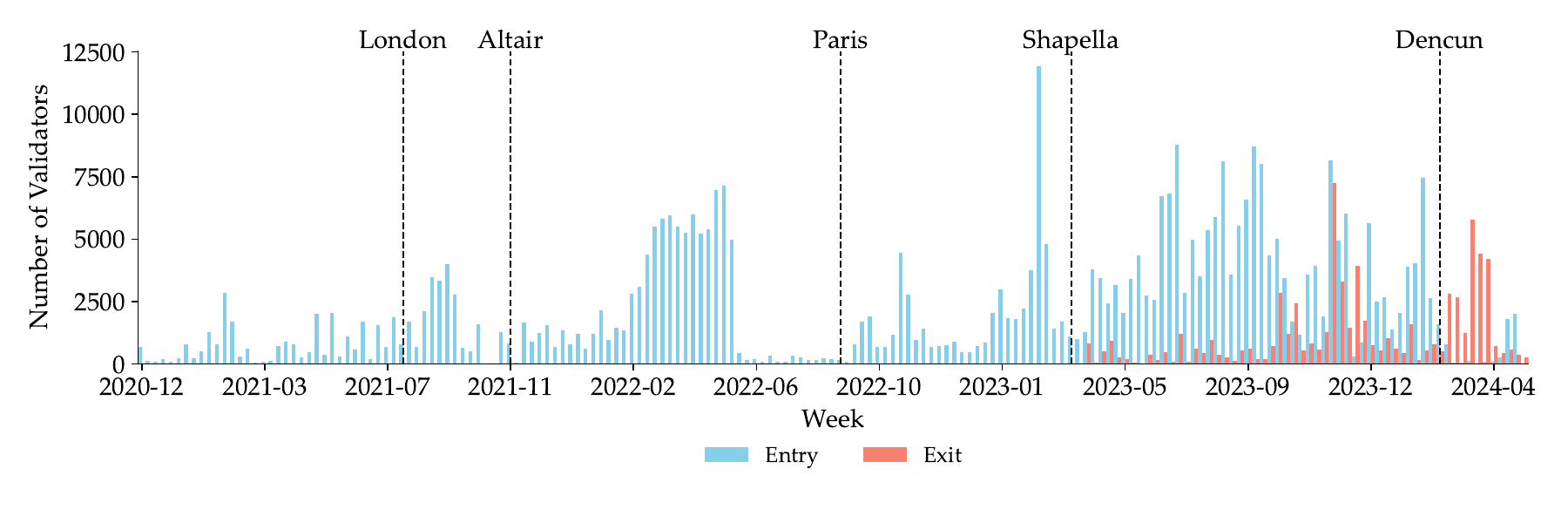}
    \end{subfigure}
    \hfill
    \caption*{\footnotesize \textit{Source}: Rated Network}
\end{figure}

To provide an overview of the Ethereum staking market, we illustrate data on the overall market shares of different staking methods\footnote{Source: Dune query by user hildobby, see \url{https://dune.com/queries/1941407/3202651}.} and on the exits and entries of validators within different staking methods\footnote{Source: Rated Network through MVI Grants Program} between 1 December 2020 and 27 May 2024. In particular, Figures \ref{fig:ETH_Staked} and \ref{fig:ETH_Staked_share} illustrate the evolution of staked ETH across different staking categories over time, and Figure \ref{fig:EntryExit} shows the entry and exit dynamics of validators across these solutions. While a portion of staking supply cannot be reliably attributed to a specific staking solution, thus introducing some degree of uncertainty to any conclusions drawn from the data, several distinct patterns emerge.

First, both centralized exchanges and liquid staking solutions have grown significantly over time, collectively capturing a substantial market share. More recently, however, the rapid rise of liquid restaking has coincided with a decline in staking via centralized exchanges and liquid staking solutions, both in absolute terms and as a share of total staked ETH. This trend suggests an element of competition between different staking methods. While large liquid staking providers benefit from strong network effects due to the issuance of LSTs, these effects seem not to have been sufficient to fully insulate them from market share erosion following the introduction of new staking innovations.

Additionally, solo staking has experienced a steady decline, both in absolute terms and as a proportion of total staking. As shown in Figure \ref{fig:EntryExit}, solo staking saw significant validator entry in the early stages of Ethereum’s PoS transition. However, this trend reversed over time, with validator exits outweighing new entries after the Shapella upgrade, which enabled stakers to withdraw their staked ETH. The Shapella upgrade also represents a noticeable shift in user behavior for staking via centralized exchanges and via liquid staking solutions. Both categories saw an increase in validator entries, while only centralized exchanges saw a sharp increase in exits. On this basis, it appears that users of centralized exchanges also valued the improved liquidity introduced by the Shapella upgrade. Of course, if this is the case, it begs the question of why users of centralized exchanges ever chose these solutions over liquid staking, which by design allowed users to retain the liquidity of their stake even before the Shapella upgrade. It may therefore be reasonable to infer that there are other user preferences that segment the staking market to a degree that has allowed centralized exchanges to retain a substantial market share even during periods when liquid staking solutions could offer far greater liquidity.

Taken together, this supports the findings of \cite{ethIssuanceDiscovery}, that the Ethereum staking market exhibits a degree of segmentation based on the type of staking solution and user preferences. While competition certainly appears to exist between liquid staking and liquid restaking—implying a shared customer base—the broader dynamics of staking supply seem to suggest that solo staking, liquid staking, and staking via centralized exchanges cater to distinct user preferences and constraints.

\section{Theoretical Model}\label{Section_Model}
The following section presents a game-theoretic framework through which we can model the strategic staking decisions of various market participants in relation to the Ethereum issuance schedule, alternative revenue sources and staking costs. In particular, the model framework enables us to analyze how differing levels of staker sophistication and SSP fee structures influence staking participation and the overall reward structure. For an overview of the parameters defined by the model, please refer to Table \ref{tab:calibration_summary}.

\subsection{Agent Types}

Building on the empirical findings in Section \ref{sec:empirical_evaluation}, the model differentiates ETH holders based on their preferences and level of technical sophistication, which in turn determines the staking options available to them. Specifically, three distinct types of ETH holders are considered:
\[
\theta \in \{\text{Retailer, Techie, Expert}\}.
\]
\noindent Each category comprises a fixed number of stakers, denoted as $N_r$, $N_t$, and $N_{ss}$ for \textit{Retailers}, \textit{Techies}, and \textit{Experts} (a category composed entirely of solo stakers as we explain below), respectively. For analytical tractability, we assume homogeneity within each group.

\subsection{Segmented Staking Market}

In addition, the model incorporates several staking methods that differ in the costs they impose on ETH holders and the revenue streams they provide. These methods include solo staking; staking through a centralized service provider (cSSP), such as a centralized exchange or a professional direct staking delegation company like Kiln; and staking through a decentralized service provider (dSSP) that provides additional DeFi yields through an LST, such as Lido. We do not model restaking explicitly.

Motivated by the market dynamics described in Section \ref{sec:empirical_evaluation} and the evidence presented by \cite{ethIssuanceDiscovery}, we model the preferences of the different agent types to depict a segmented staking market.\footnote{Segmented staking markets are also considered in other studies, such as \citet{artem2024maximum}.} That is, each ETH holder type selects a single staking method, and does not diversify across options. Specifically:
\begin{itemize}
    \item Assuming limited technical ability and a preference for ease of use, retailers either stake with a cSSP or abstain, which prevents them from engaging in solo staking or using a dSSP.
    \item Techies prefer dSSPs over cSSPs because they receive higher returns from LSTs. That is, we model revenue and cost functions such that Techies' profits are larger than those of Retailers. As such, techies are assumed to value liquidity and favor decentralized intermediaries that provide additional yields through liquid staking tokens (LSTs).
    \item Experts are assumed to exclusively choose to solo stake—despite potentially lower profits—due to an intrinsic (but unmodeled) utility derived from contributing to decentralization. As a consequence, Experts solo stake or leave the staking market.
\end{itemize}

\noindent In sum, each ETH holder allocates their entire stake to the option that yields the highest utility for their type. This framework emphasizes clear segmentation rather than hybrid strategies.\footnote{An alternative approach could involve portfolio management. However, this kind of approach makes more sense when the different asset risk-adjusted returns have low or negative correlation. This might not be the case when considering only different kind of staking methods as in our model.}

\subsection{Staking Revenues}

Depending on their choice of staking option, ETH holders may receive revenues from three different sources: 1) consensus layer revenue, 2) MEV revenue, and 3) additional DeFi yield.

The first is \textbf{consensus layer revenue}, determined by the protocol’s issuance schedule. The annual consensus yield provided to an ETH holder when staking is denoted as $y_i(D)$. Under the current issuance schedule, the annual consensus yield is given by

\begin{equation*}
y_i(D) = \frac{cF}{\sqrt{D}} = \frac{2.6 \cdot 64}{\sqrt{D}},
\end{equation*}

\noindent where $D$ represents the total amount of ETH staked in the protocol. In the following, we will essentially compare model outcomes under the current issuance schedule $y_i(D)$ to alternative specifications $y_i'(D)$.

The second revenue source is \textbf{MEV revenue}. Validators selected as proposers in block production can extract additional revenue from structuring transactions, referred to in the following as MEV revenue. Letting $N = D/32$ denote the total number of validators, the probability that an ETH holder controlling $d_i/32$ validators is chosen as a proposer is $d_i/D$. Conversely, if a staker joins a staking pool (i.e. stakes with an intermediary), proposers may share MEV revenues with attesters. The probability that the proposer belongs to the pool is $d_{pool}/D$, and a validator in the pool with stake $d_i$ receives a share of $d_i/d_{pool}$. Thus, the expected MEV revenue for a staker in a pool remains the same as for a solo staker. In both cases the expected annual MEV revenue $y_v$ for an ETH holder is theoretically given by

\begin{equation*}
y_v \cdot \frac{d_i}{D}.
\end{equation*}

The third source of revenue is \textbf{additional DeFi yield}. ETH holders utilizing a dSSP may earn additional yields by reinvesting the LST they receive. The annual yield from this source of revenue is simply defined as $y_d$.

\subsection{Staking Costs}

From the perspective of the ETH holder, staking is also associated with costs. In the present model, the cost function faced by an ETH holder is allowed to differ for each staking solution. We define the cost function to include fixed costs and non-linear variable costs:
\begin{equation*}
C(d) = C+c\,d^{\alpha}.
\end{equation*}

Later, we calibrate the model looking for the set of parameters that gives a similar distribution of staking supply than the one observed in the market at the moment of this study. We also assume that the costs functions are such that there are interior solutions.\footnote{This assumption is standard in models where agents face no binding constraints on their choices, such as budget constraints.} The assumptions made when calibrating the cost parameters are described in Section \ref{sec:calibration}.

Furthermore, ETH holders staking through an intermediary incur \textbf{fees}, which are modeled as a tax on revenues and as such treated separately from the cost functions. The cSSP fee rate is denoted as $f_c$, and the dSSP fee rate as $f_d$, both of which are assumed to be exogenous in the baseline model. As such, fees are simply assumed to be fixed and not determined by the strategic behavior of the staking service providers in the market. Subsequent extensions will relax this assumption and to some extent allow fees to be determined as a function of the profit maximization problem of the staking service providers in the model. Again, the assumptions made when calibrating the fee parameters are described in Section \ref{sec:calibration}. 

\subsection{Agent Staking Decisions}

Each ETH holder strategically decides how much to stake in order to maximize their profits, taking into account the staking behavior of others. That is, stakers anticipate the actions of other participants and their effects on the inflationary reward, and adjust their own staking levels accordingly to achieve a profit-maximizing outcome. The resulting model solution thus represents a Nash-equilibrium. We refer to Section \ref{sec:derivation} of the Appendix for a detailed derivation.

The following model can be interpreted as one in which a staker is endowed with a fixed amount of fiat currency and must decide how much to allocate to ETH staking versus how much to consume or use for other purposes. To simplify the analysis, we introduce several assumptions. The opportunity cost of foregone consumption is implicitly captured by a convex cost function.\footnote{This also justifies the use of a convex cost function, as it reflects the increasing marginal cost of substituting away from alternative consumption or uses.} As a result, the staker's utility (or profit) function is concave, leading to interior solutions to the maximization problem—even in the absence of binding budget constraints, as we assume here.

For simplicity, we abstract from fluctuations in the ETH price, including expectations about future price movements and any associated capital gains or losses. That is, we assume a constant ETH price, thereby isolating the staking decision from market volatility and the negative effects of inflation on the token price. In addition, stakers are assumed to not discount the future. Under all these assumptions, the staker’s problem reduces to maximizing expected annual profit. We next describe each staker type decision problem:

Given the deposits of other stakers $\{d_{ss},d_r,d_t\}$, a representative \textit{Retailer} chooses a deposit $\hat{d}_{r}\geq 0$ to maximize
\begin{align}
\Pi_r\equiv&\max_{\hat{d}_{r}}\, \Big\{ (1-f_c)\cdot\Big[y_i(D)\cdot\hat{d}_r + y_v\cdot P_r(D)\Big] - \Big(C_{r}+c_{r}\cdot\hat{d}_r^{\alpha_{r}}\Big) \Big\},
\label{eq:retailer}\\
&\text{such that:} \nonumber\\
& \Pi_r\geq 0 \nonumber \\
& D=N_{ss}\cdot d_{ss}+N_t\cdot d_t+\hat{d}_r+(N_r-1)\cdot d_r \in[0,120M], \nonumber \\
& P_r(D)=\frac{\hat{d}_{r}}{N_{ss}\cdot d_{ss}+N_t\cdot d_t+\hat{d}_r+(N_r-1)\cdot d_r}\in[0,1], \nonumber
\end{align}
where $f_c$ is the fee charged by $cSSPs$.

Let $\hat{d}_{r}^*$ denote the optimal choice of the maximization problem. Due to homogeneity within each staker type category, all Retailers must choose the same deposit level in equilibrium. That is, in equilibrium, the individual optimal choice $\hat{d}_{r}^*$ satisfies $\hat{d}_{r}^* = d_{r}$.

$P_{r}(D)$ denotes the probability that a staker with deposit $\hat{d}_{r}$ becomes a proposer. We model this probability as the ratio of the staker’s deposits to the total deposits, as indicated in the denominator. A more accurate representation would model this as the ratio of the number of validators controlled by the staker to the total number of validators. This would require a step function that increments with every 32 ETH, corresponding to the validator activation threshold. For tractability, we instead model $P_{r}$ as a continuous function of $\hat{d}_{r}$, which simplifies the analysis.

Similarly, each \textit{Techie} chooses a deposit $\hat{d}_{t}\geq 0$ to maximize
\begin{align}
\Pi_t\equiv&\max_{\hat{d}_{t}}\, \Big\{ (1-f_d)\cdot\Big[y_i(D)\cdot\hat{d}_t + y_v\cdot P_t(D)\Big] + y_d\cdot \hat{d}_t - \Big(C_{t}+c_{t}\cdot\hat{d}_t^{\alpha_{t}}\Big) \Big\},
\label{eq:techie}\\
&\text{such that:} \nonumber \\
& \Pi_t\geq\Pi_r, \nonumber \\
& D=N_{ss}\cdot d_{ss}+\hat{d}_t+(N_t-1)\cdot d_t+N_r \cdot d_r \in[0,120M], \nonumber \\
& P_t(D)=\frac{\hat{d}_{t}}{N_{ss}\cdot d_{ss}+\hat{d}_t+(N_t-1)\cdot d_t+N_r \cdot d_r}\in[0,1]. \nonumber
\end{align}

Homogeneity again implies $\hat{d}_t^*=d_t$. Notice that this type of staker pays a fee $f_d$ and enjoys of DeFi yields $y_d$.

A representative \textit{Expert} chooses a deposit $\hat{d}_{ss}\geq 0$ to maximize
\begin{align}
\Pi_{ss}\equiv&\max_{\hat{d}_{ss}}\, \Big\{ y_i(D)\cdot \hat{d}_{ss} + y_v\cdot P_{ss}(D) - \Big(C_{ss}+c_{ss}\cdot\hat{d}_{ss}^{\alpha_{ss}}\Big) \Big\},
\label{eq:solo_staker}\\
&\text{such that:}\nonumber \\
& \Pi_{ss}\geq 0 \nonumber \\
& D=\hat{d}_{ss}+(N_{ss}-1)\cdot d_{ss}+N_r\cdot d_r+N_{t}\cdot d_t\in[0,120M] \nonumber \\
& P_{ss}(D)=\frac{\hat{d}_{ss}}{\hat{d}_{ss}+(N_{ss}-1)\cdot d_{ss}+N_r\cdot d_r+N_{t}\cdot d_t} \in [0,1]. \nonumber
\end{align}

Assuming homogeneity, all solo stakers choose the same deposit in equilibrium, i.e., $\hat{d}_{ss}^*=d_{ss}$.

\subsection{Model Extensions}\label{model_extension}

We extend the baseline game-theoretic model by incorporating key refinements to the characteristics of staking agents, building each extension sequentially. First, we introduce solo stakers facing greater variance in MEV revenues than other stakers. Next, we relax the assumption that all stakers are fully strategic by introducing inattentive agents who do not respond to changes in the issuance curve. Finally, we enhance the model of the intermediary market by incorporating a dSSP with market power. 

\subsubsection{Variability in MEV rewards for solo stakers }\label{extention_mev}

In the baseline model, expected MEV revenues are assumed to be identical regardless of whether an ETH holder chooses to solo stake or joins a pool via a staking service provider. However, since MEV revenues can vary substantially over time, this formulation does not account for the fact that solo stakers may require several years to propose a block with high MEV, unlike other stakers in a pool who can benefit from smoothing MEV revenues over time. To incorporate such difference, we introduce the variance over such reward (denoted below as $Y_{MEV}$) for a moderately risk-averse solo stakers, while we assume that for the other types that variance can be neglected, i.e., pools are large enough. Hence, the maximization problem of a solo staker (assumed to be of type  \textit{Expert}) becomes:\footnote{See the Appendix for the derivation and discussion.}

\begin{equation}\label{equation_SS_problem_with_variance}
         \max_{\hat{d}_e} \quad y_i(D) \cdot \hat{d}_{e}+[y_v\cdot P_{ss}(D)- y_v^2\cdot P_{ss}(D)^2]-(C_{ss}+c_{ss}\cdot \hat{d}_{e}^{\alpha_{e}}),
\end{equation}
when we use
\begin{align*}
    E[Y_{MEV}]-Var[Y_{MEV}]&=y_v\cdot P_{ss}(D)- y_v^2\cdot P_{ss}(D)^2.
\end{align*}

\subsubsection{Inattentive Agents}\label{extention_inattentive}

Further, we divide the \textit{Retailer} category into two subgroups: (1) institutions and (2) inattentive retail investors. The latter group is assumed to be unresponsive to changes in the issuance yield. This behavior could be interpreted as the agents being inattentive to changes in issuance or that their profit-maximizing decisions result in a corner solution, as is the case in the theoretical results obtained by \cite{julianmasolo}.\footnote{Evidence from other blockchains suggests that the presence of inattentive stakers may be a relevant phenomenon. For instance, \citet{Liu_et_al2024Video_CompetitionInCryptoStaking} find that approximately $80\%$ of delegator accounts on the Cardano network are "sleepy" delegators, representing around $50\%$ of the total staked volume. An interesting research direction is to study how prevalent inattentive staking behavior is within the Ethereum ecosystem. For our analysis, we assume that only $20\%$ of the volume staked belongs to inattentive stakers.} The primary purpose of incorporating this type of agent into the model is to assess how their presence influences the staking behavior of other types.

\subsubsection{Intermediary dSSP with Market Power}\label{extention_market_power}

So far, we have simplified the role of decentralized staking service providers (dSSPs) in the market. In particular, the above model specifications implicitly assume that dSSPs operate under perfect competition and that they provide the staking service themselves. 

\begin{figure}[H]
    \centering
    \caption{dSSP as Intermediaries between ETH holders and cSSPs}
    \label{fig:Lidointermediary}
    \includegraphics[width=0.7\linewidth]{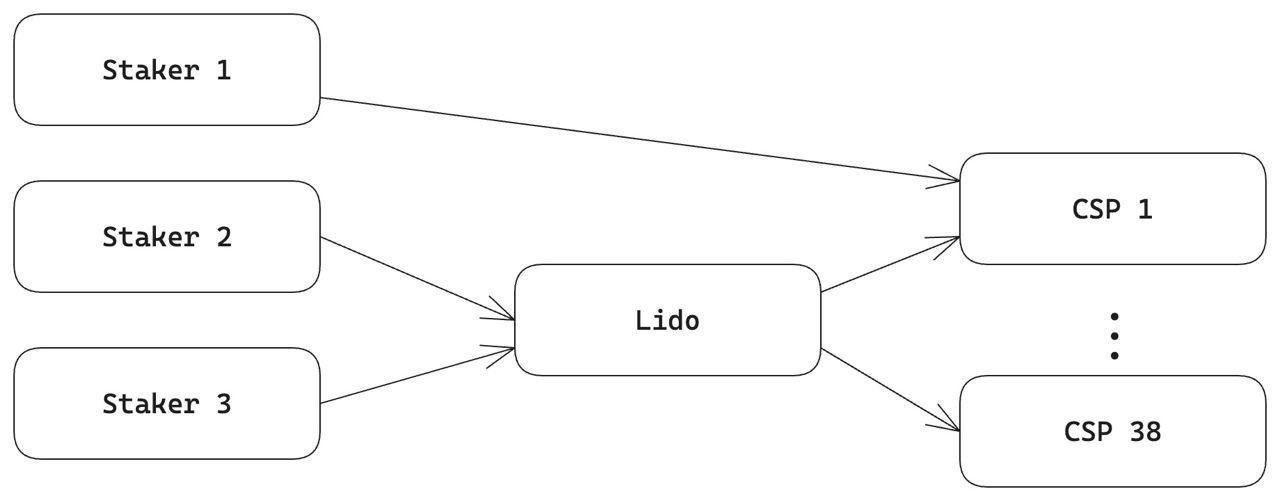}
\end{figure}

A more realistic approach would be to model dSSPs as intermediaries or middleware between stakers and centralized staking service providers (cSSPs). For example, if an ETH holder decides to stake with Lido, the stake is managed by a cSSP. Lido itself does not directly provide the staking service, but rather facilitates the connection between the ETH holder and the cSSP. In this arrangement, the ETH holder benefits from Lido by receiving a derivative representing its stake. Figure \ref{fig:Lidointermediary} illustrates this setup in a simplified way.

In the following, we extend our model to incorporate the interactions between an intermediary, stakers, and multiple centralized staking service providers (cSSPs). Our framework specifically captures Lido, the dominant player in this intermediation space. Rather than assuming perfect competition, we model the decentralized staking service provider (dSSP) as a player with market power, building on previous extensions that account for inattentive retail stakers and the higher variance in execution rewards faced by solo stakers.

Our objective is to examine how changes in the issuance curve influence the dSSP’s fee-setting decisions and, consequently, the overall staking level. We preserve the timing structure from earlier model iterations: the dSSP sets its fees first, followed by stakers determining their deposit levels.

In the following, we describe the updated profit maximizing behaviors of the players shown in Figure \ref{fig:Lidointermediary}: the intermediary dSSP (Lido), its potential users (ETH holders of type \textit{Techie}), and the cSSPs. The strategic behavior of the other actors, in particular the ETH holder types \textit{Expert} and \textit{Retailer}, remains unchanged.

\begin{itemize}
    \item \textbf{Techies:} We maintain the assumption of a segmented staking market and assume that the ETH holder type \textit{Techie} chooses the dSSP as an intermediary. In doing so, they are indifferent to the specifics of the dSSP-cSSP relationship; their decision is based solely on Lido's fees ($f_d$), their expected rewards, and associated costs. Essentially, they solve the same optimization problem as in equation (\ref{eq:techie}): each Techie selects $\hat{d}_t\geq 0$ to maximize their profit
    \begin{equation}
              \max_{\hat{d}_{t}}\quad (1-f_d) \cdot \left[ y_i(D) \cdot \hat{d}_{t}+y_v\cdot  P_{t}(D) \right]+y_{d}\cdot \hat{d}_{t}- (C_t+c_{t}\cdot \hat{d}_t^{\alpha_{t}}),
    \end{equation}
    \item \textbf{Intermediary (Lido):} To maximize its profits, the dSSP sets both the fee $f_d$ charged to Techies and the fee $\hat{f}_d$ paid to cSSPs. The dSSP holds the bargaining power to determine these fees, as there are many cSSPs competing to be integrated into the platform. We assume that the dSSP charges a uniform fee to each cSSP and distributes stakes evenly among them. The profit maximization problem of the dSSP is therefore:
    \begin{align}\label{equation_Lido_Profit_Function}
    	\max_{f_d,\hat{f}_d}&\quad (f_d-\hat{f}_d)\cdot\left[y(D_{-t},D_t(f_d))\cdot D_t(f_d)+y_{v}(D_{-t},D_t(f_d)) \right]-C_{I}(D_t(f_d)),\\
    	& \text{s.t. }\pi_{cSSP}\geq 0 \nonumber \tag{PC}
    \end{align}
    where $D_t=\sum \hat{d}_t$ denotes the total stakes by \textit{Techies}, and $D_{-t}$ the stakes from non-\textit{Techies}. We assume that the dSSP enjoys economies of scale, i.e., $C_I(D_t)$ is a concave function. In setting $\hat{f}_d$, the dSSP ensures that cSSP profits, $\pi_{cSSP}$, remain at or above zero. In other words, the dSSP considers the cSSPs' participation constraints (PC) to secure sufficient cSSP participation for allocating \textit{Techies}' stakes.
    \item \textbf{cSSP:} Consider any cSSP $m=1...M$. This cSSP sets the fee $f_c^m$ charged to stakers who bypass the dSSP and stake directly with it (e.g., see Staker 1 in Figure \ref{fig:Lidointermediary}).

    \begin{align}\label{equation_cSSP_Profit_Function_LidoCase}
    	\max_{f_c^m}&\quad f_c^m\cdot\left[y(D)\cdot D_i^m+y_{v}(D_i^m,D) \right]+\hat{f}_d\cdot\left[y(D)\cdot D_t^m+y_{v}(D_t^m,D) \right] \\
        &\text{ }-C_c(D_t^m+D_i^m), \nonumber
    \end{align}
    
    where $D_i^m$ represents the total stakes from institutions that choose to stake with $m$. As previously mentioned, we assume that Lido distributes $D_t$ uniformly across all the cSSPs:
        
    \begin{equation*}
    	D_t^m=\frac{D_t}{M}.
    \end{equation*}
    
    Again, we assume that cSSPs benefit from economies of scale, i.e., $C_c(D_t)$ is a concave function.
\end{itemize}

\subsection{Results}\label{Result}

In the following, we compare the equilibrium outcomes of the game-theoretic model under two different issuance schedules. The first issuance schedule corresponds to the one currently implemented by the Ethereum protocol:  
\begin{equation*}
    y_i(D) = \frac{2.6 \cdot 64}{\sqrt{D}},
\end{equation*}
where $D$ represents the staking deposit level. The second issuance schedule follows one of the alternative proposals presented by \cite{elowsson2024tempered}:  
\begin{equation*}
    y'_i(D) = \frac{2.6\cdot 64}{\sqrt{D} \cdot (1+k\cdot D)},
\end{equation*}
where $k=2^{-25}$ and $D$ again represents the total amount of ETH staked. 

\subsubsection{Numerical Simulation}\label{simulation}

As a first step, the general relationships between the model parameters and the resulting equilibrium outcomes are explored. The goal is to develop an intuition for the relevant factors driving equilibrium staking decisions and how they drive differences in the outcomes under the two consensus issuance schedules. This is done by running simulations of the model, randomly varying the parameter values in each simulation run.\footnote{To account for the large parameter space, we conduct 5 million simulation runs for each of the two issuance schedules.}

\begin{figure}[H]
\caption{Relationship between cost parameters and change in equilibrium staking decision}
\centering
\label{fig:costs_simulations}
\begin{subfigure}[t]{0.3\textwidth}  
    \centering
    \includegraphics[width=\linewidth]{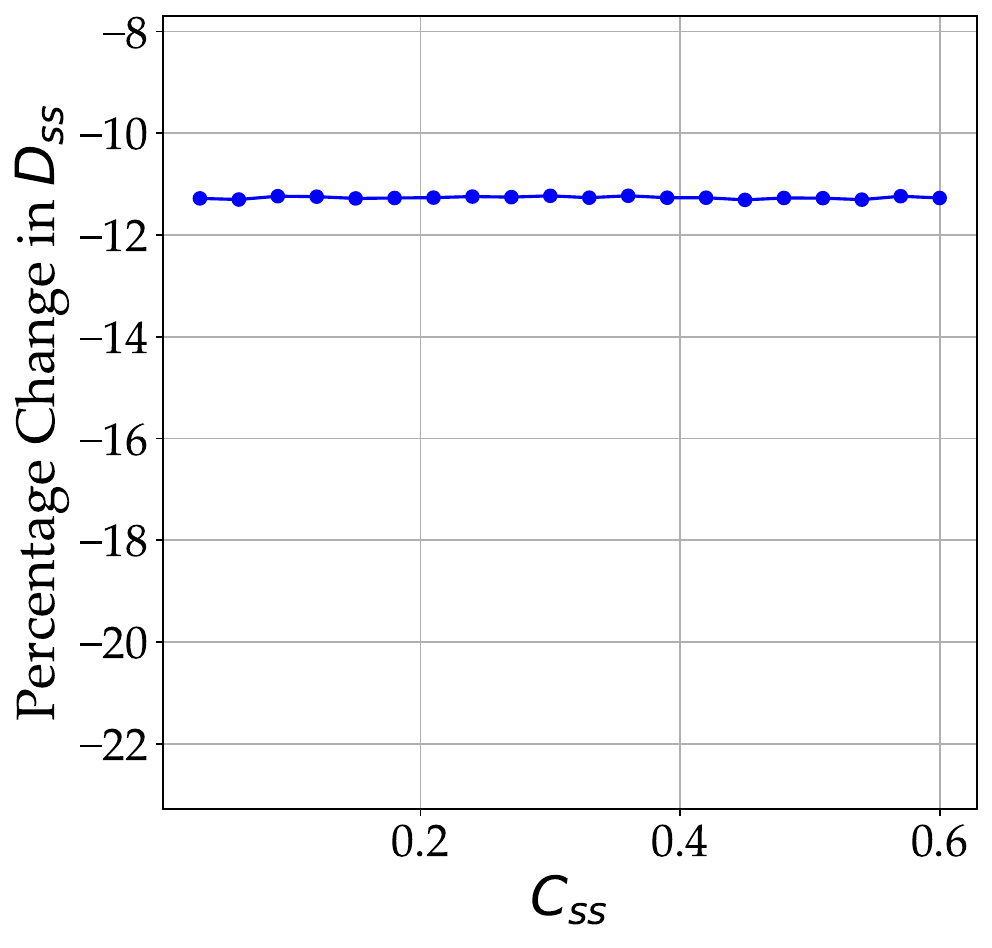}
    \caption{$D_{ss}$ vs. $C_{ss}$}
    \label{fig:1}
\end{subfigure}%
\hfill
\begin{subfigure}[t]{0.3\textwidth}  
    \centering
    \includegraphics[width=\linewidth]{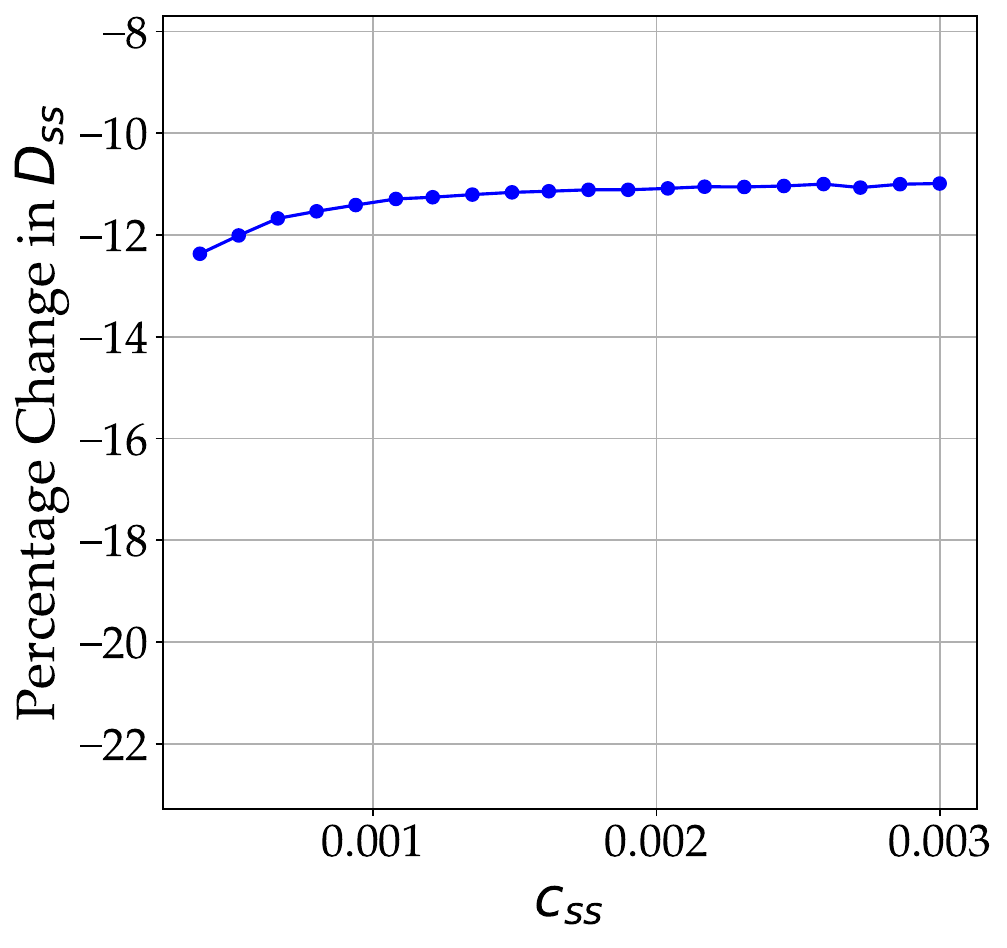}
    \caption{$D_{ss}$ vs. $c_{ss}$}
    \label{fig:2}
\end{subfigure}%
\hfill
\begin{subfigure}[t]{0.3\textwidth}  
    \centering
    \includegraphics[width=\linewidth]{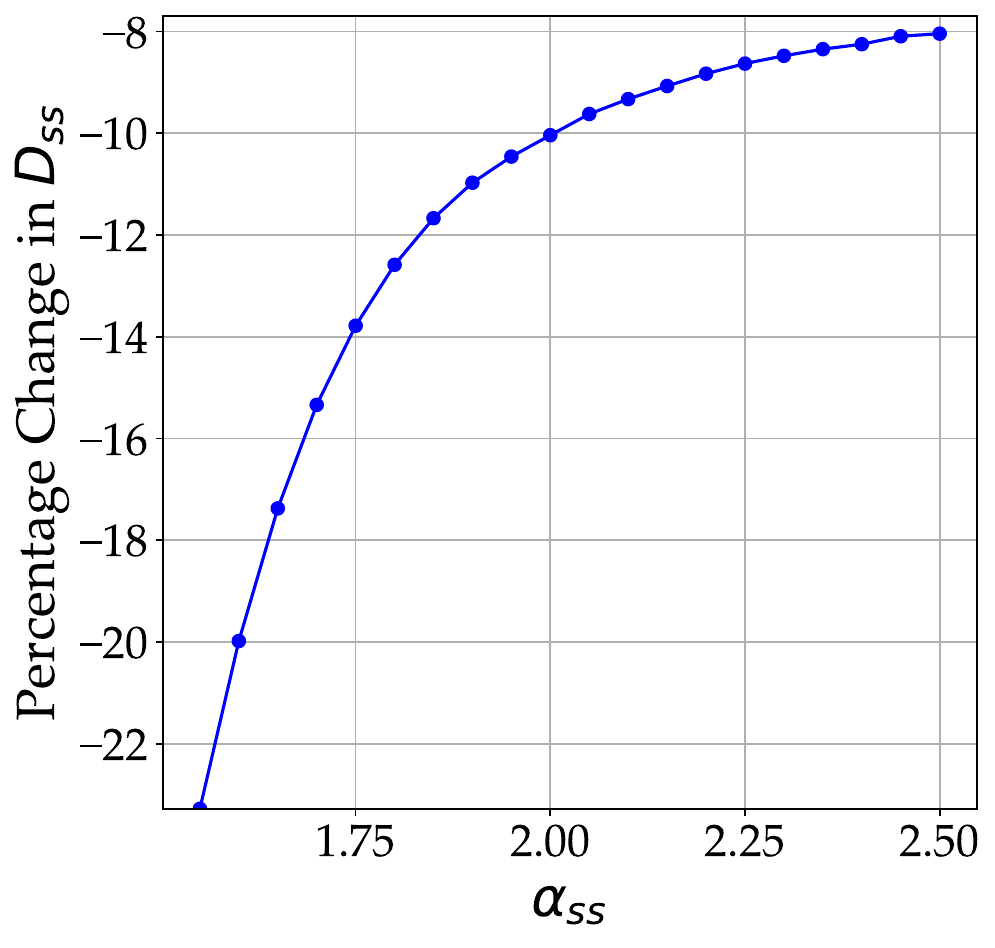}
    \caption{$D_{ss}$ vs. $\alpha_{ss}$}
    \label{fig:3}
\end{subfigure}%
\hfill
\begin{subfigure}[t]{0.3\textwidth}  
    \centering
    \includegraphics[width=\linewidth]{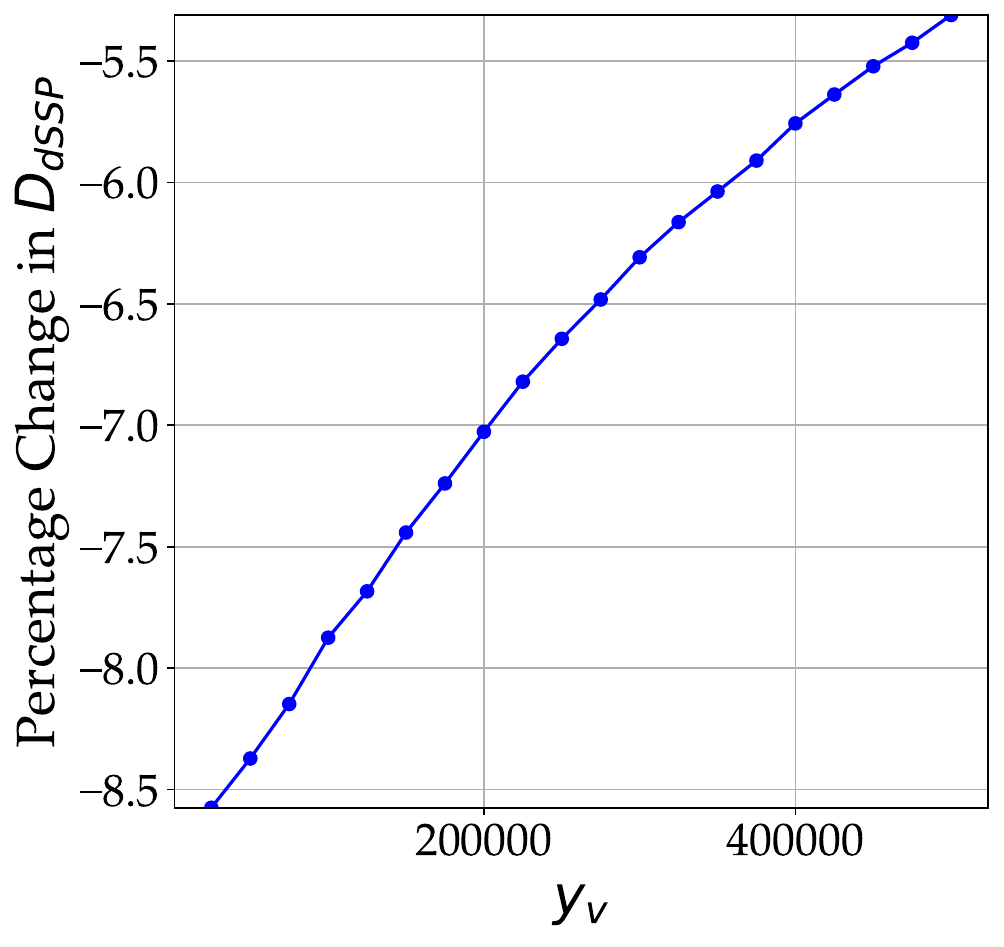}
    \caption{$D_{dSSP}$ vs. $y_{v}$}
    \label{fig:4}
\end{subfigure}%
\hfill
\begin{subfigure}[t]{0.3\textwidth}  
    \centering
    \includegraphics[width=\linewidth]{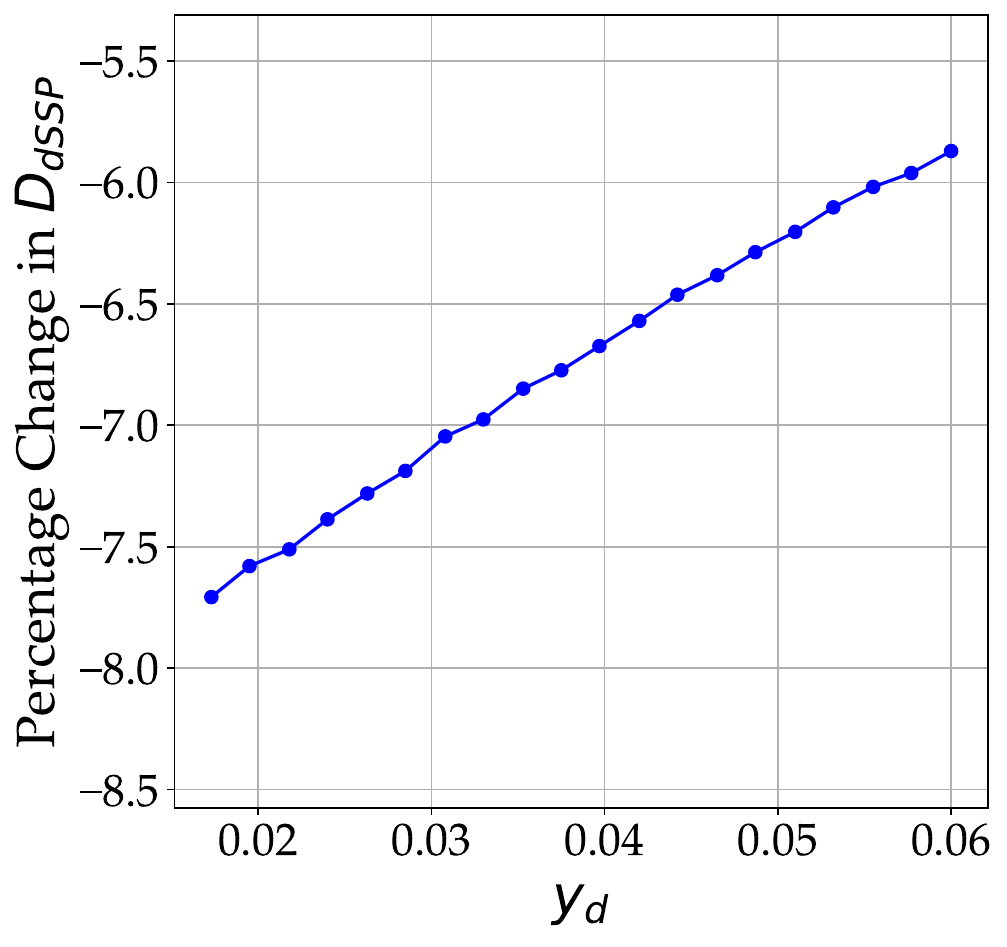}
    \caption{$D_{dSSP}$ vs. $y_{d}$}
    \label{fig:5}
\end{subfigure}%
\hfill
\begin{subfigure}[t]{0.3\textwidth}  
    \centering
    \includegraphics[width=\linewidth]{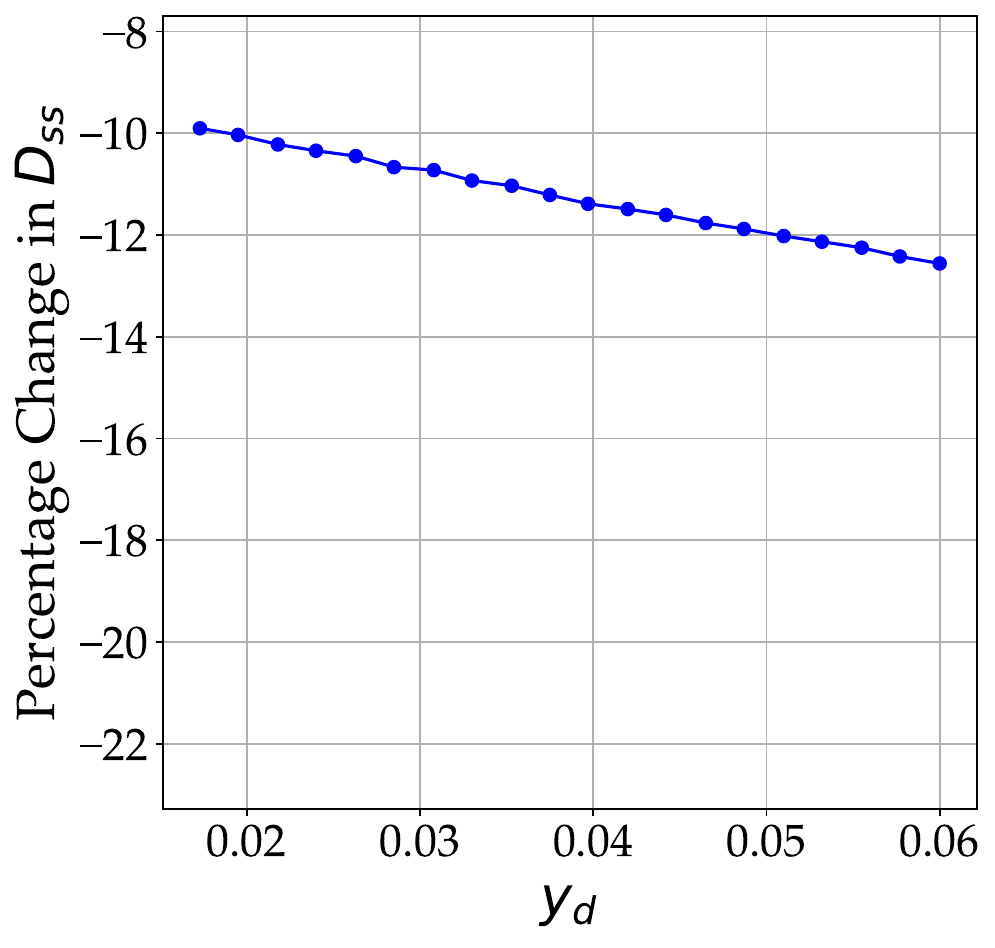}
    \caption{$D_{ss}$ vs. $y_{d}$}
    \label{fig:6}
\end{subfigure}%
\caption*{\footnotesize \textit{Note}: The plots display the binned average percentage change in the total staking supply of different staking methods when transitioning from $y_i(D)$ to $y'_i(D)$, categorized by different levels of model parameters. The number of bins is set to 20.}
\end{figure}

The simulations yield several key observations that offer insight into how different model parameters influence changes in equilibrium outcomes when transitioning from $y_i(D)$ to $y'_i(D)$:

\begin{itemize}
    \item \textbf{Observation 1:} Everything else held equal, a staking solution having higher marginal costs is associated with smaller adjustments in the equilibrium staking supply when transitioning from $y_i(D)$ to $y'_i(D)$. This effect is illustrated in Figures \ref{fig:1} - \ref{fig:3}, using solo staking as an example.
    
    \item \textbf{Observation 2:} Stakers having access to additional MEV revenues or DeFi yields is associated with a weaker response in their equilibrium staking supply when transitioning from $y_i(D)$ to $y'_i(D)$. Figures \ref{fig:4} and \ref{fig:5} illustrate this effect using staking via dSSP as an example.

    \item \textbf{Observation 3:} The sensitivity of one type of agent to changes in the issuance schedule affects the sensitivity of other agent types to changes in the issuance schedule. For instance, when \textit{Techies} become less sensitive to issuance schedule changes—such as when additional DeFi yields are high—the sensitivity of \textit{Experts} to a change in the issuance schedule increases. Figure \ref{fig:6} illustrates this effect.
\end{itemize}

\subsubsection{Model Calibration}\label{sec:calibration}

\begin{figure}[H]
\caption{Median distance (L2 norm) of simulated market shares compared to observed market shares for different values of $c_t$ and $c_r$}
\centering
\label{fig:simulation_l2_norm}
\includegraphics[width=0.6\linewidth]{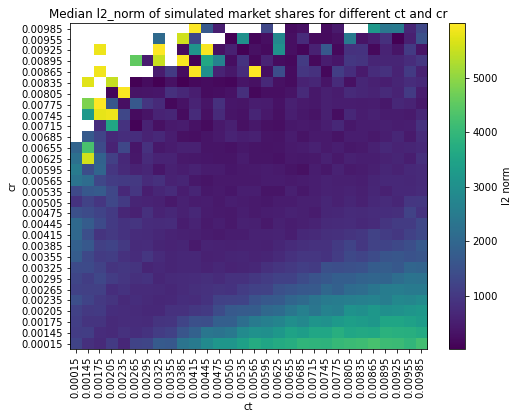}
\caption*{\footnotesize \textit{Note}: This plot displays the median distance (L2 norm) between simulated market shares and observed market shares for different values of $c_t$ and $c_r$.}
\end{figure}

We calibrate the baseline model with specific parameters settings and compare Nash equilibria under two issuance schedules. Given the number of parameters to be calibrated, we find that estimating model parameters based on a method of simulated moments, i.e., algorithmically setting parameters to minimize the distance between simulated and observed market shares, does not provide a stable best fit. In other words, we find that a variety of possible parameter combinations can yield simulated market shares that are consistent with the observed data. This finding is illustrated in Figure \ref{fig:simulation_l2_norm}. For different combinations of parameter settings for $c_t$ and $c_r$, we find that a range of parameter settings are able to minimize the distance between simulated market shares and observed market shares. This issue is exacerbated when allowing the algorithm to search freely over an increasing number of parameters. 

As an alternative approach, we propose parameter settings that are broadly consistent with the available literature and our qualitative understanding of the Ethereum staking market and that in the baseline model, achieve a close fit to the observed market shares (see Table \ref{tab:calibration_distribution}). Additionally, we conducted a survey on the Reddit forum \textit{ethstaker}, focusing on the cost structure and decisions faced by solo stakers.\footnote{See \url{https://www.reddit.com/r/ethstaker/comments/1dneiy3/solo_stakers_questionnaire/}}

\begin{table}[H]
\centering
\caption{Reference staking distribution used in parameter calibration }
\label{tab:calibration_distribution}
\begin{tabular}{p{6cm} >{\raggedleft\arraybackslash}p{2cm} >{\raggedleft\arraybackslash}p{2cm}}
    \toprule
    \textbf{Staking Method} & \textbf{M ETH} & \textbf{\%} \\
    \midrule
     dSSP & $18$ & $54.1\%$\\
     cSSP & $14.4$ & $44.6\%$\\
     Solo Staking & $0.9$ & $2.7\%$\\
    \midrule
     TOTAL & $34.2$ & \\
    \bottomrule
\end{tabular}
\caption*{\footnotesize Note: The estimated number of validators is based on a simplification of the categorization reported by \citep{artem2024maximum} and is calculated as the amount of staked ETH divided by 32. We categorize DSMs, LRTs, Whales, and Unidentified as dSSPs. Meanwhile, cSSPs consist of CSPs and CeXs.}
\end{table}

First, we assume that solo staking incurs fixed costs, while staking via a dSSP or cSSP does not:
\[
C_{ss} = 0.4,\quad C_{t} = 0,\quad C_{r} = 0.
\]

Contrary to staking through an intermediary, solo staking requires an individual to set up an Ethereum validator node. To do this, solo stakers can purchase dedicated hardware, subscribe to a cloud computing service, or use existing hardware. The costs associated with each of these options vary substantially. To further complicate matters, a range of hardware specifications can be used to run a node, each with varying costs. To date, the literature has made highly simplifying assumptions about the hardware costs of solo validators. In particular, both \cite{pa7x1} and \cite{artem2024maximum} assume that a solo staker factors in 1000 USD to initially purchase the necessary hardware, which they amortize over a period of 5 years. The anecdotal evidence collected from the survey on \textit{ethstaker} generally supports this assumption but also reveals substantial variation owing to the fact that some stakers use existing hardware and do not purchase dedicated hardware for staking, i.e. likely treat their hardware expenditure as sunk costs.

Similar to hardware costs, we assume costs for energy and internet to be fixed, since they are a fundamental requirement to running an Ethereum node but do not scale when adding additional validators to a node. Again, \cite{pa7x1} and \cite{artem2024maximum} assume around 1.4 to 2 USD in energy costs per week and around 0 to 12 USD per week for internet costs. These assumptions are generally supported by the anecdotal evidence from the \textit{ethstaker} survey.

Taken together, we assume fixed costs of $\approx 1'000 \$/year$. This includes hardware which is usually assumed to be amortized in $5$ years ($200-400$ $\$/year$), high-speed and stable internet connection ($50$ $\$/month$), and additional electricity expenditures ($100$ $\$/year$). Denoting this cost in terms of ETH, at a price of ETH in USD of 2'400 USD, we obtain a parameter setting of $C_{ss} = 0.4$. Again, we assume that staking through an intermediary does not create any fixed costs from the perspective of the ETH holder. Therefore, we set $C_{r} = 0$ and $C_{r} = 0$.

Further, we are argue the following general assumptions to hold regarding variable costs
\[
c_{ss} < c_{t} < c_{r} \quad \text{and} \quad \alpha_{ss} > \alpha_{t} = \alpha_{r}.
\]

The literature is quite sparse on variable staking costs. As a result, we rely primarily on anecdotal evidence to support these assumptions. We argue that imposing the above conditions may be reasonable when considering the maintenance costs of solo staking. As the number of validators increases, complexity and operational overhead may increase non-linearly. For example, more validators require better network management. Managing and updating multiple nodes, as well as handling security patches and upgrades, requires more effort as the infrastructure scales. Based on the results of the \textit{ethstaker} survey, it is also plausible to assume that, at least at lower levels of staking, stakers are hobbyists and do not price in maintenance and upkeep costs. As staking operations increase in size, we expect that individual stakers begin to price in maintenance and upkeep costs when choosing their optimal.

Moreover, we assume that the variable costs for staking via a cSSP exceed those for staking via a dSSP. This assumption is based on the observation that, in addition to the inherent risks of staking (e.g., slashing risks), centralized exchanges introduce additional custodial risks—risks that are absent in decentralized staking service providers such as Lido, which do not take custody of users' assets. In addition, institutional stakers staking via a centralized solution may accrue additional legal costs. Finally, we assume that $\alpha_{j} > 1$ for all staking methods $j$. This allows for convex cost functions which capture the fact that ETH holders may be resource constrained. This assumption is standard in models where agents face no binding constraints on their choices, such as budget constraints.

Overall, due to the argumentative nature of our calibration approach, we argue that the results presented below should not be interpreted as predictive point estimates. Rather, the focus should be on the relative changes and dynamics observed between the Nash equilibria under the two issuance schedules being considered.

\begin{table}[H]
\centering
\caption{Summary of parameter settings} 
\label{tab:calibration_summary}
\resizebox{\textwidth}{!}{%
\begin{tabular}{p{2cm} p{10cm} p{3.5cm}} 
     \toprule
    \textbf{Parameter} & \textbf{Description} & \textbf{Parameter Value} \\
    \midrule
    $N_{ss}$& Number of agents of type \textit{Expert} & $N_{ss} = 25'000$\\
    \midrule
    $N_t$ & Number of agents of type \textit{Techie} & $N_t = 200'000$\\
    \midrule
    $N_r$ & Number of agents of type \textit{Retailer} & $N_r = 925'000$\\
    \midrule
    $y_{v}$ & Annual MEV revenue & $y_{v} = 300'000$ \\
    \midrule
    $y_{d}$ & Annual DeFi yield & $y_{d} = 0.02$\\
    \midrule
    $f_d$ & Fee dSSP & $f_d = 0.1$ \\
    \midrule
    $f_c$ & Fee cSSP & $f_c = 0.25$ \\
    \midrule
    $C_{ss}$ & Fixed cost parameter solo staking & $C_{ss} = 0.4$ \\
    \midrule
    $c_{ss}$ & Variable cost parameter solo staking & $c_{ss} = 0.00053 $ \\
    \midrule
    $\alpha_{ss}$ & Exponent cost function solo staking & $\alpha_{ss} = 2$ \\
    \midrule
     $C_{t}$ & Fixed cost parameter for \textit{Techie} staking via dSSP & $C_{t} = 0 $ \\
    \midrule
    $c_{t}$ & Variable cost parameter for \textit{Techie} staking via dSSP & $c_{t} = 0.0038 $ \\
    \midrule
    $\alpha_{t}$ & Exponent cost function for \textit{Techie} staking via dSSP & $\alpha_{t} = 1.5$ \\
    \midrule
     $C_{r}$ & Fixed cost parameter for \textit{Retailer} staking via cSSP & $C_{r} = 0 $ \\
    \midrule
    $c_{r}$ & Variable cost parameter for \textit{Retailer} staking via cSSP & $c_{r} = 0.0048$ \\
    \midrule
    $\alpha_{r}$ & Exponent cost function for \textit{Retailer} staking via cSSP & $\alpha_{r} = 1.5$ \\
    \bottomrule
\end{tabular}
}
\end{table}

\newpage
\subsubsection{Comparative Analysis}

Below we analyze the equilibrium results of the game-theoretic model based on specific parameter calibrations described in the previous section. Effectively, we compare the equilibrium staking outcomes of a calibrated baseline model with the equilibrium outcomes of a calibrated model under the alternative issuance schedule $y'_i(D)$.

\vspace{0.5cm}
\noindent \textbf{Baseline Model}

\noindent Table \ref{tab:staking_summary_baseline} presents the total amount of ETH staked and the corresponding profits for each type of ETH holder under various issuance schedules, as well as the relative changes in these values resulting from the change in the issuance schedule. 

\begin{table}[H]
\centering
\caption{Nash equilibria of baseline model}
\label{tab:staking_summary_baseline}
\resizebox{\textwidth}{!}{%
\begin{tabular}{l l r r r r r r r r r}
\toprule
    & & \multicolumn{3}{c}{\textbf{$y_i(D) = \frac{2.6 \cdot 64}{\sqrt{D}}$}} & \multicolumn{3}{c}{\textbf{$y'_i(D) =\frac{2.6 \cdot 64}{\sqrt{D} \cdot (1 + k \cdot D)}$}} & \multicolumn{2}{c}{\textbf{$\Delta$}} \\
\cmidrule(lr){3-5} \cmidrule(lr){6-8} \cmidrule(lr){9-10}
    \textit{Types} & \textit{Num} & \textit{Deps} & \textit{Ratio} & \textit{Profit$^a$} & \textit{Deps} & \textit{Ratio} & \textit{Profit$^a$} & \textit{Deps} & \textit{Profit} \\
\midrule
    e & $25K$ & $0.9M$& $2.7\%$& $0.8\%$& $0.74M$& $2.97\%$& $0.2\%$& $-17.8\%$&$-75\%$\\ 
\midrule
    t & $200K$ & $18M$& $54.1\%$& $1.8\%$& $14.3M$& $57.5\%$& $1.6\%$& $-20.6\%$& $-11.1\%$\\ 
\midrule
    r & $925K$ & $14.4M$& $43.2\%$& $0.95\%$& $9.8M$& $39.53\%$& $0.78\%$& $-31.5\%$& $-17.9\%$\\ 
\midrule
 \textbf{Total} & & \textbf{$33.3M$} & & & \textbf{$24.84M$} & & & \textbf{$-25.4\%$} &\\
\bottomrule
\multicolumn{10}{l}{\footnotesize{$^a$ Profit per ETH staked}} \\
\end{tabular}
}
\end{table}

\noindent The first observation is that a change in the issuance schedule results in an approximate 25\% reduction in the total amount of staked ETH. The relative decrease is most pronounced for the ETH holder type \textit{Retailer}, followed by \textit{Techie} and \textit{Expert}. Consequently, although the overall staking supply declines across all types, the market share of solo staking and staking via dSSPs increases. Based on the intuitions developed in Section \ref{simulation}, this result can be explained by \textit{Experts} facing high marginal costs when solo staking, making them less sensitive to changes in the issuance schedule, as well as \textit{Techies} having access to additional external DeFi yields, again making them less sensitive to changes in the issuance schedule. Despite a substantial reduction in their equilibrium staking supply, all ETH holder types experience lower staking profits per ETH staked under the adjusted issuance schedule. This decline is particularly pronounced for the \textit{Expert} type, who uniquely must cover their fixed costs and higher average costs with a reduced revenue base.

\vspace{0.5cm}

\newpage
\noindent \textbf{Expected MEV Revenues}

\noindent Next, we repeat the above analysis for the extended version of the model, where the type \textit{Expert} is assumed to be risk-averse and as such discounts future MEV revenues. Table \ref{tab:staking_summary_MEV} again reports the resulting equilibrium outcomes under the two issuance schedules.

\begin{table}[H]
\centering
\caption{Nash equilibria of extended model with discounted MEV revenues}
\label{tab:staking_summary_MEV}
\resizebox{\textwidth}{!}{%
\begin{tabular}{l l r r r r r r r r r}
\toprule
    & & \multicolumn{3}{c}{\textbf{$y_i(D) = \frac{2.6 \cdot 64}{\sqrt{D}}$}} & \multicolumn{3}{c}{\textbf{$y'_i(D) =\frac{2.6 \cdot 64}{\sqrt{D} \cdot (1 + k \cdot D)}$}} & \multicolumn{2}{c}{\textbf{$\Delta$}} \\
\cmidrule(lr){3-5} \cmidrule(lr){6-8} \cmidrule(lr){9-10}
    \textit{Types} & \textit{Num} & \textit{Deps} & \textit{Ratio} & \textit{Profit$^a$} & \textit{Deps} & \textit{Ratio} & \textit{Profit$^a$} & \textit{Deps} & \textit{Profit} \\
\midrule
    e & $25K$ & $0.77M$& $2.3\%$& $0.6\%$& $0.58M$& $2.3\%$& $0.0\%$& $-24.6\%$&$-100\%$\\ 
\midrule
    t & $200K$ & $18M$& $54.3\%$& $1.8\%$& $14.3M$& $57.8\%$& $1.6\%$& $-20.6\%$& $-11.1\%$\\ 
\midrule
    r & $925K$ & $14.4M$& $43.4\%$& $0.95\%$& $9.9M$& $39.85\%$& $0.78\%$& $-31.25\%$& $-17.9\%$\\ 
\midrule
 Total & & $33.2M$ & & & $24.76M$ & & & $-25.4\%$&\\
\bottomrule
\multicolumn{10}{l}{\footnotesize{$^a$ Profit per ETH staked}} \\
\end{tabular}
}
\end{table}

\noindent When incorporating the variance in MEV rewards for a risk-averse and small solo staker of type \textit{Expert}, their relative reduction in staking supply becomes more pronounced compared to the baseline model. As outlined in Section \ref{simulation}, this outcome is driven by the following dynamic: if agents expect to have access to lower expected MEV revenues, they become more sensitive to changes in the issuance schedule. As a result, a change in the issuance schedule no longer leads to an increase in the market share of solo stakers. At the same time the staking outcomes for the remain staking solutions and ETH holder types remain virtually unchanged relative to the baseline model. Since \textit{Expert} solo stakers represent only a small fraction of the overall population, their heightened sensitivity to issuance changes has little impact on \textit{Techies} and \textit{Retailers}. Additionally, the negative effect of issuance reductions on the expected profits per staked ETH for solo stakers is further exacerbated, with expected profits even turning negative.

\newpage

\noindent \textbf{Inattentive Agents}

\noindent Now, we incorporate to the model with the higher variability in MEV rewards for solo stakers, the differentiation of the ETH holder type \textit{Retailer} into (1) institutions and (2) an inattentive retail investor, as described in Section \ref{extention_inattentive}. Table \ref{tab:staking_summary_extended} reports the corresponding results. 

\begin{table}[H]
\centering
\caption{Nash equilibria of extended model with inattentive agents}
\label{tab:staking_summary_extended}
\resizebox{\textwidth}{!}{%
\begin{tabular}{l l r r r r r r r r r}
\toprule
    & & \multicolumn{3}{c}{\textbf{$y_i(D) = \frac{2.6 \cdot 64}{\sqrt{D}}$}} & \multicolumn{3}{c}{\textbf{$y'_i(D) =\frac{2.6 \cdot 64}{\sqrt{D} \cdot (1 + k \cdot D)}$}} & \multicolumn{2}{c}{\textbf{$\Delta$}} \\
\cmidrule(lr){3-5} \cmidrule(lr){6-8} \cmidrule(lr){9-10}
    \textit{Types} & \textit{Num} & \textit{Deps} & \textit{Ratio} & \textit{Profit$^a$} & \textit{Deps} & \textit{Ratio} & \textit{Profit$^a$} & \textit{Deps} & \textit{Profit} \\
    \midrule
    e & $25K$ & $0.78M$& $2.4\%$& $0.6\%$& $0.57M$& $2.2\%$& $0.0\%$& $-26.9\%$ & $-100\%$\\ 
\midrule
    t & $200K$ & $18.1M$& $54.7\%$& $1.8\%$& $13.7M$& $53.3\%$& $1.6\%$& $-24.3\%$& $-11.1\%$\\ 
\midrule
    i & $300K$ & $7.7M$& $23.3\%$& $0.95\%$& $4.9M$& $19.1\%$& $0.76\%$& $-36.4\%$& $-20.0\%$\\ 
\midrule
    r & NA & $6.5M$& $19.6\%$& NA & $6.5M$ & $25.3\%$ & NA & $0\%$ & NA \\    
\midrule
 Total & & $33.1M$ & & & $25.7M$ & & & $-22.4\%$&\\
\bottomrule
\multicolumn{10}{l}{\footnotesize{$^a$ Profit per ETH staked}} \\
\end{tabular}
}
\end{table}

\noindent In the presence of the inattentive ETH holder type—whose staking supply remains constant across both issuance schedules—the relative reduction in the overall staking supply is somewhat lower than in the baseline model. Meanwhile, the ETH holder types that do adjust their staking supply in response to the issuance schedule experience a decline in their market shares under the adjusted schedule. Again this is consistent with the intuitions developed in Section \ref{simulation}. When one ETH holder type does not modify its staking supply in response to an issuance reduction, the adjustments made by the other types tend to be more pronounced. The solo staking type \textit{Expert} still experiences a larger reduction in their staking supply than the \textit{Techie} type, who continues to benefit from external DeFi yields, making their equilibrium staking supply less sensitive to changes in protocol issuance. Consequently, the decline in staking profits per ETH staked is even greater for the \textit{Expert} type.

\vspace{0.5cm}

\noindent \textbf{Intermediary dSSP with Market Power}

\noindent Finally, we conduct a comparative analysis of equilibrium outcomes under the assumption of an intermediary dSSP, similar to Lido, that acts as a platform between ETH holders and SSPs. Importantly, the fee set by the dSSP is endogenous to the model and therefore subject to the strategic response of the dSSP to a change in the issuance schedule. In solving the model, we make some additional assumptions and simplifications relative to the model presented in Section \ref{extention_market_power}:

\begin{itemize}
    \item We assume that, as a result of change in the consensus issuance schedule, the dSSP (Lido) may change $f_d$ but not $\hat{f}_d$.
    \item To simplify our analysis, we do not explicitly solve the dSSP's full profit maximization problem as outlined above, since our primary focus is on determining the direction in which the fee should move. To solve exactly for the optimal fee, we would first need to calibrate Lido's cost function. Instead, given a cost structure with economies of scale, we calculate Lido's profits under the current issuance schedule and the fee assumed in the baseline model calibration. Then, under the updated issuance yield curve, we simply identify the new fee as the fee that would yield a similar profit, and then use that fee to derive the new staking equilibrium.  
    \item Similarly, we do not solve for the optimal choice of $f_c^m$ for the cSSPs, and when comparing equilibrium outcomes under different issuance schedules, we simply consider the variable to be fixed to the same calibration as in the baseline model. This may cause $f_c^m$ to be set too high, since in the baseline version of the model, cSSPs are assumed to be more like a CeX, and in the current extension, they are more akin to SSPs like P2P and Figment. However, this assumption ensures a better comparability with the previous results and we are therefore better able to isolate the effect of the market power of the dSSPs on the equilibrium results.
\end{itemize}

\noindent Table \ref{tab:staking_summary_extended_market_power} reports the results of the comparative analysis under these assumptions, in particular the fees set by the dSSP and the distribution of stakes across different ETH holder types and staking methods. 

\begin{table}[H]
\centering
\caption{Nash equilibria of extended model with intermediary dSSP}
\label{tab:staking_summary_extended_market_power}
\resizebox{0.7\textwidth}{!}{%
\begin{tabular}{l l r r r r r r r r}
\toprule
    & & \multicolumn{2}{c}{\textbf{$y_i(D);f_d=10\%$}} & \multicolumn{2}{c}{\textbf{$y'_i(D);f_d=13\%$}} & \multicolumn{1}{c}{\textbf{$\Delta$}} \\
\cmidrule(lr){3-4} \cmidrule(lr){5-6} \cmidrule(lr){7-7}
    \textit{Types} & \textit{Num} & \textit{Deps} & \textit{Ratio} & \textit{Deps} & \textit{Ratio} & \textit{Deps} \\
    \midrule
    ss & $25K$ & $0.78M$& $2.4\%$ 
    & $0.57M$ & $2.2\%$ 
    & $-26.9\%$ 
    \\ 
    \midrule
    t & $200K$ & $18M$ & $54.5\%$ 
    & $13.4M$ & $52.5\%$ 
    &  $-25.5\%$ 
    \\ 
    \midrule
    i & $300K$ & $7.7M$& $23.3\%$ 
    & $5M$ & $19.6\%$ 
    & $-35\%$ 
    \\ 
    \midrule
    r & NA & $6.5M$& $19.7\%$ 
    & $6.5M$ & $25.5\%$ 
    & $0\%$ 
    \\    
    \midrule
     Total & & $33M$  & & $25.5M$ &  & $-22.7\%$\\
    \bottomrule
\end{tabular}
}
\end{table}

The adjustment of the issuance schedule is associated with an increase of $30\%$ in $f_d$. As a result, the decrease in the total amount staked through the dSSP is greater and the staking distribution becomes more concentrated among the cSSPs compared to the previous model iterations where the dSSP fees did not adjust in response to a change in the issuance schedule. Put differently, the increase in $f_d$ primarily impacts stakes from \textit{Techies}, who reduce their participation. Institutions, in turn, benefit from this reduced Techie participation, increasing their staking level by $2\%$—approximately equivalent to the decrease in Techie participation. Since cSSPs are the providers for Institutional stakes, this shift could be interpreted as an increase in concentration within the staking market. 

\section{Empirical Analysis}\label{empirical analysis}

In the following, we employ an instrumental variables (IV) approach to obtain unbiased estimates of the slope of the supply curves across different staking methods.  

\subsection{Estimation Approach}

\begin{figure}[h!]
\caption{Illustration of Instrumental Variable Approach}
\label{Figure_IV_summary}
	\centering\includegraphics[width=16cm]{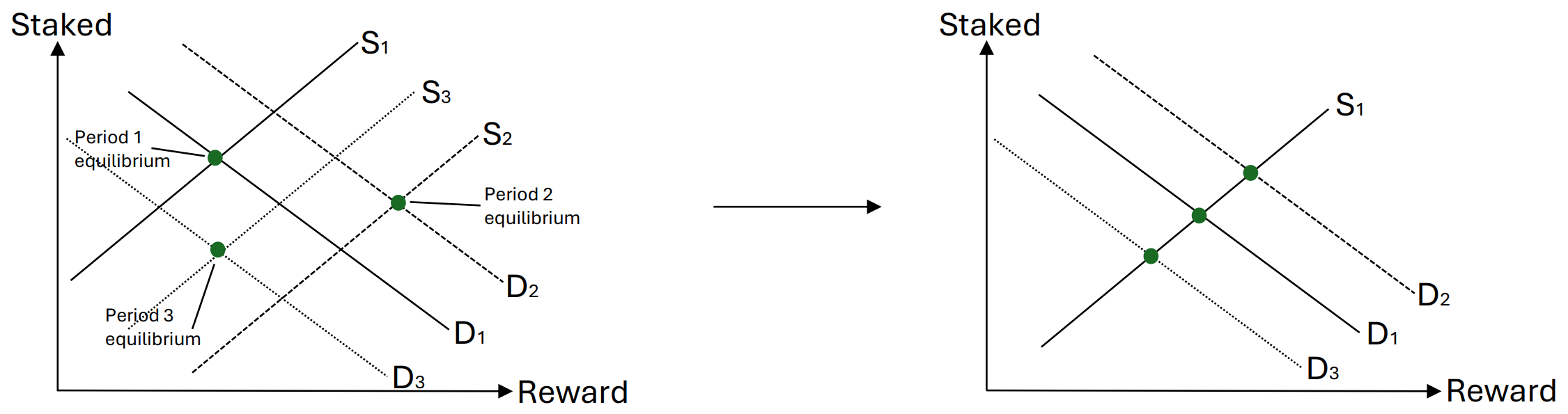}
\end{figure}

To address endogeneity concerns in the estimates of staking supply elasticity, we employ an instrumental variable (IV) approach. Endogeneity arises in this context because any observed combination of staking yields and staking amounts reflects a market equilibrium where supply and demand for staking meet. Changes in this equilibrium can result from shifts in either the supply or demand curve. As depicted in Figure  \ref{Figure_IV_summary}, the IV approach addresses this issue by identifying instruments that shift the demand for staked ETH without simultaneously affecting the factors that determine the supply curve. This allows us to isolate the slope of the supply curve at the equilibrium point, i.e. the yield elasticity of supply. 

\newpage 

\noindent Consider the following linear model of the staking supply curve: 

\begin{equation} 
D_{jt} = \beta_0 + \beta_1 y_{jt} + \gamma' W_t + \epsilon_{jt}, 
\end{equation} 

\noindent where $D_{jt}$ is the the staking supply at time $t$ for staking method $j$, $y_{jt}$ is the real staking yield at at time $t$ for staking method $j$, and $W_t$ a vector of control variables. To obtain a consistent estimate of $\beta_1$, we introduce an instrumental variable $Z_{jt}$ that shifts $y_{jt}$ independently of $\epsilon_{jt}$. A valid instrument $Z_{jt}$ must satisfy two key conditions:

\begin{enumerate}
    \item Instrument relevance ($\mathrm{Cov}(Z_{jt}, y_{jt})\neq 0$): The instrument must identify meaningful variation in the independent variable of interest $y_{jt}$. In the literature, this condition is typically argued to be fulfilled if the F-Statistic of the first stage regression, i.e. the regression of $y_{jt}$ on $Z_{jt}$ is greater than 10.
    \item Instrument exogeneity ($\mathrm{Cov}(Z_{jt}, \epsilon_{jt}) = 0$): As outlined above, the instrument should be exogenous and the effect of the instrument $Z_{jt}$ on $D_{jt}$ should run solely through its effect on $y_{jt}$. In general, this condition cannot be tested empirically.
\end{enumerate}

\noindent Under these conditions, variations in $Z_i$ induce exogenous variation in $y_t$, which can be used to identify $\beta_1$. Formally, we implement a two-stage least squares (2SLS). In the first stage, we regress the endogenous regressor on the instrument: 

\begin{equation} 
y_t = \pi_0 + \pi_1 Z_t + \gamma' W_t + \nu_i, \quad \text{(First Stage)} \end{equation} 

\noindent obtaining the fitted values $\hat{X}_i$. In the second stage, we regress $Y_i$ on the fitted values from the first stage: 

\begin{equation} 
D_t = \beta_0 + \beta_1 \hat{y}_t + \gamma' W_t + \epsilon_i, \quad \text{(Second Stage)} 
\end{equation} 

\noindent yielding the 2SLS estimate $\hat{\beta}_1^{IV}$. In the following, we discuss EIP updates and gas fees as potential valid instrumental variables.



\subsubsection{EIPs as Instrument}

Staked ETH is the way through which Ethereum secures its Proof-of-Stake (PoS) consensus, for which the Ethereum blockchain rewards stakers with ETH. Put differently, \textit{demand for staked ETH is the protocol's demand for security}. With this in mind, Ethereum Improvement Proposals (EIPs) involving upgrades that affect how much stakers earn for providing PoS security arguably represent shifts in the demand curve from which one could trace out the staking supply curve. 

Regarding the condition of \textit{instrument exogeneity}, it is not unambiguous that EIP changes do not at all affect the supply curve. For example, part of the Shapella upgrade on April 12th 2023 was introducing the ability for stakers to withdraw their rewards and original stakes. Until then, all rewards were locked on the consensus layer (Beacon Chain). To the extent that future rewards are discounted (due to the time value of money, etc.) the ability to withdraw contemporaneously increases ETH rewards from the protocol, shifting the demand curve to the right. However, it could be argued that would-be stakers anticipating this could be more willing to supply a certain amount of ETH for every yield level, that changes in their expectations would shift the supply curve down/to the right. 

Further, The EIP upgrades were all announced long before they occurred, and being mainnet forks, necessarily permanent. Thus, it could be argued that shifts in the supply curve would have been spread out over time, such that the effects captured by a dummy variable for the day of the fork would have captured much more of the direct effect on demand. 

A dummy variable was created for each of six relevant EIP forks, starting with the London upgrade on August 5th 2021 through to March 13th 2024's Dencun upgrade. Changes in PoS rewards captured may be in real or nominal ETH, e.g., EIP-1559 introduced through the London fork changed the fee market structure, separating transaction fees into base and priority fees, the former of which was burned. This effectively increased real ETH rewards for staking and represents the demand curve for staked ETH shifting right. Formally, we estimate the following first stage regression:

\begin{equation} \label{eqn}
y_t = \pi_0 + \pi_1 Bellatrix_t + \pi_2 Paris_t + \pi_3 Shapella_t + \pi_4 Dencun_t + \pi' W_t + \nu_i,
\end{equation}

\noindent where $y_t$ represents the log of annualized staking rewards in USD. In the second-stage regression, the predicted value of log USD staking rewards $\hat{y}$ is then used as an explanatory variable:

\begin{equation} \label{eqn}
D_t = \gamma_0 + \gamma_1\hat{y}_t + \gamma' W_t + u_t.
\end{equation}

\noindent where $D_t$ is the natural log of total quantity staked at time $t$. To control for other factors influencing changes in staking supply decisions, we include additional control variables, e.g., a proxy for the risk-free rate, as well as a dummy for ETH price crashes, to control for the opportunity cost and psychological cost of staking, respectively.

\subsubsection{Gas Fees as Instrument}

In addition, we explore gas fees as a potential instrument. The causal pathway from gas fees to staking rewards -- and subsequently to staking supply -- is well-defined and straightforward. Gas fees, on one hand, directly influence the rewards that validators earn from staking. As gas fees increase due to, for example, higher network activity, the effective yield from staking rises. This makes gas fees a relevant instrument, as these fees clearly impact the variable of interest -- the staking rewards. On the other hand, gas fees are primarily driven by the demand for and value of transactions on the Ethereum network, which is influenced by factors like network usage, DeFi activity, NFT trading, or other dApp interactions. These factors are largely independent of ETH holders’ staking supply decision, making gas fees an exogenous variable in relation to the staking supply. Thus, using gas fees as an instrument leverages a key component of the Ethereum ecosystem that naturally fluctuates due to external factors (such as user demand) while directly influencing staking rewards but not directly impacting staking supply, making it a robust choice for this analysis. Additionally, the regular fluctuations and public availability of gas fees data enable high-frequency and reliable analysis. 

We thus employ the 2SLS approach outline above using gas fees as the instrument. By log-transforming the variables, we can interpret the estimated coefficient $\hat{\beta}_1^{IV}$ as the supply elasticity of staking with respect to staking rewards, denominated in USD. Formally, we again estimate the following first stage regression

\begin{equation} \label{eqn}
log(y_{t}) = \pi_0 + \pi_1 + log(r_{t}) + \nu_i,
\end{equation}

\noindent where $r_t$ represent total gas fees paid in the protocol at time $t-1$. We then specify the second-stage regression as

\begin{equation} \label{eqn}
log(D_t) = \gamma_0 + \gamma_1log(\hat{y}_{t}) + \gamma' W_t + u_t.
\end{equation}

\subsection{Data}

Two main data sets are employed for this analysis. The first is from Rated, an independent provider of node and node operator data, with supplementary data from Dune and Renzo analytics which were kindly shared by fellow grantees. This was the longest set of validator data the study had access to, stemming from the launch of Beacon Chain (i.e., Day 0 or December 1st 2020) through to June 2024. Additional data, e.g., for different PoS L1 real rewards, as well as annualized USD returns from ETH staking, were extracted from Staking Rewards's API. For ETH, the earliest rewards data was naturally from 1st December 2020. Daily-level data on gas fees were obtained via Dune Analytics. Natural logs of the variables were used, per the derivations above.

\subsection{Results}
\subsubsection{EIPs as an IV}

As previously outlined, for an instrument to be valid, it needs to be both \textit{relevant} to the endogenous variable and \textit{uncorrelated} with the unobserved error term. While verifying the latter condition is not straightforward, the former is easily tested with an OLS regression of the endogenous variable - the natural log of dollar rewards to staking. With an F-statistic of 428.2, the instruments prove to be collectively relevant to variations in dollar rewards. To the extent the above argument is a compelling one, combined with the encouraging 'pseudo' first-stage regression, one could implement the EIPs-2SLS method. 

\begin{table}[H]
\centering
\caption{Testing the relevance of instruments}
\label{tab:Testing the relevance of instruments} 
\tiny  
\setlength\tabcolsep{14pt}  
\renewcommand{\arraystretch}{1.5}
\begin{tabular}{l d{1.3} d{1.3}} \\ \hline
& \multicolumn{1}{c}{OLS} \\ \hline
Bellatrix  &  0.0136 \\
 & (0.101) \\
Paris & 0.1339 \\
 & (0.095) \\ 
Shapella & 0.7645^{***} \\
 & (0.025) \\
Dencun & 0.5601^{***} \\
 & (0.041) \\
Constant  & 20.6109^{***} \\
 & (0.040) \\ \hline
Observations  & \multicolumn{1}{c}{423}\\
F-statistic & \multicolumn{1}{c}{428.2} \\
R-squared   & 0.804 \\ \hline
\multicolumn{2}{c}{ Standard errors in parentheses} \\
\multicolumn{2}{c}{ *** p$<$0.01, ** p$<$0.05, * p$<$0.1} \\
\end{tabular}
\end{table}
From Table \ref{tab:2SLS with Dollar Rewards and Event Dummies}, it can be seen that compared with the OLS case, the coefficient on natural log of dollar rewards from 2SLS is \textit{higher}, respectively at 0.2086 and 0.4640. Both estimates are \textit{positive} and \textit{significant}. This result is non-trivial, as without controlling for the effect of supply shifts, to the extent observed staked amount and yield combinations reflect supply changes more than demand changes (i.e., trace out the demand curve via supply shifts), the coefficient would be negative. Furthermore, the F-statistic of the first-stage regression (i.e., natural log of dollar rewards on the EIP dummies) is 428.2, indicating strong relevance.


The 2SLS estimate above suggests that for every percentage increase in dollar staking reward, staked amount increases by 0.46 percent. The higher value from 2SLS suggests that not accounting for supply-side shocks, the OLS estimator \textit{underestimates} the dollar reward-elasticity of staking supply. One possible reason for this is that the supply curve shifted to the left during the period of our study. There are numerous possible reasons for this, such as the advent of higher-yielding PoS alternatives motivating substitution away from Ethereum. 

Focusing on the remaining 2SLS estimates, the coefficient on the risk-free rate (approximated by the three-month dollar treasury yield) was also positive and significant at the 5\% level, at 0.0714. This is in line with the Capital Asset Pricing Model, which states that owners of capital (e.g., ETH) need to be compensated for taking on more perceived risk. Interestingly, neither the coefficient on the ETH price crash dummy and nor that of FTX collapse is significant (0.0186 and 0.0349, respectively).

Examining the results for solo stakers, the null hypothesis of no correlation could not be rejected for the dollar rewards rate (-0.0308), though, again the coefficient on risk free rate was positive and actually \textit{higher} than for the overall population (0.1015 vs. 0.0714). Again the coefficient on the FTX collapse was not significant. More interestingly, the effect from ETH's flash crash was highly negative and significant (-0.1434) at 0.1\%, indicating that while the overall population did not discernibly react to the crash, solo stakers did, reducing the amount staked by over 14\%.

A possible way of interpreting these results is that solos are motivated by a sense of altruism and are not generally swayed by small changes in dollar rewards, yet due to their limited ability to benefit from economies of scale and leverage, they generally have a higher minimal rewards threshold than non-solo stakers, below which solo stakers exit the market. Simply put, the prospect of markedly reduced future earnings causes solos to react more stronger than the overall population would, hurting decentralization. This is in line with the theoretical model's finding that solos are more sensitive to changes in rewards than other types of stakers.

\begin{table}[H]
\centering
\caption{2SLS with event dummies}
\label{tab:2SLS with Dollar Rewards and Event Dummies} 
\tiny  
\setlength\tabcolsep{14pt}  
\renewcommand{\arraystretch}{1.5}
\begin{tabular}{l d{1.3} d{1.3} d{1.3} d{1.3}} \hline
& \multicolumn{1}{c}{OLS} & \multicolumn{1}{c}{2SLS} &  \multicolumn{1}{c}{2SLS Solos} \\ \hline
ETH Rewards  & 0.2086^{***} & 0.4640^{***} &  -0.0308 \\
 & (0.023) & (0.042)  &  (0.019)\\
ETH Flash Crash & 0.1753^{***} & 0.0186 & -0.1434^{***} \\
 & (0.021) & (0.031)  &  (0.014)\\ 
FTX Collapse & -0.1085^{***} & 0.0349 & -0.0246 \\
 & (0.033) & (0.042) & (0.019)\\
Risk-free rate & 0.2190^{***} &  0.0714{**} & 0.1015^{***}\\
 & (0.020) & (0.029) & (0.013)\\
Constant  & 7.9750^{***} & 3.1535^{***} & 10.1725^{***} \\
 & (0.443) & (0.795)  & (0.358) \\ \hline
Observations  & \multicolumn{1}{c}{423} & \multicolumn{1}{c}{423} & \multicolumn{1}{c}{423}\\
F-statistic & \multicolumn{1}{c}{655.4} & \multicolumn{1}{c}{524.5} &  \multicolumn{1}{c}{189.0} \\
R-squared   & 0.862 & 0.823 & 0.649 \\ \hline
\multicolumn{4}{c}{ Standard errors in parentheses} \\
\multicolumn{4}{c}{ *** p$<$0.01, ** p$<$0.05, * p$<$0.1} \\
\end{tabular}
\end{table}

\subsubsection{Gas fees as an IV}

In our second estimation approach, we perform separate regressions for solo staking and for total staking, with results shown in Tables \ref{tab:solo_staking_daily} and \ref{tab:total_daily}, respectively. There are two notable insights from these results. In the first-stage regressions, gas fees exhibit a positive and statistically significant effect on staking rewards, with F-statistics exceeding the threshold of 10, thereby meeting the relevance condition for instruments. This strong association reinforces the appropriateness of gas fees as an instrumental variable for staking rewards. In the second stage, the estimated elasticity of staking supply with respect to staking rewards is 1.184 for solo stakers alone and 1.078 for all stakers, both statistically significant at the 1 percent level. This result suggests that, consistent with \cite{Eloranta}, solo stakers are slightly more responsive to changes in staking rewards than the overall staking cohort.

We validate the robustness of these findings through additional tests. Following a similar approach to \cite{cong2022tokenomics}, we incorporate various lags for both staking rewards and gas fees, observing minimal variation in results, which supports the stability of our estimates. Furthermore, when aggregating the data to a weekly frequency (see Tables \ref{tab:solo_staking_weekly} and \ref{tab:total_weekly} in the Appendix), the results remain consistent, underscoring the reliability of our elasticity estimates.

Again, the results are broadly consistent with the results of the theoretical model, which suggest that solo stakers are more sensitive to changes in the issuance yield compared to other staking categories. Still, these estimates are based on very short-term variations in gas fees. As such, it may be difficult to extrapolate the estimation results to an equilibrium model, such as the one developed above. 

\begin{table}[H]
    \centering
    \caption{Solo Staking, Daily Data}
    \label{tab:solo_staking_daily} 
    \tiny  
    \setlength\tabcolsep{4pt}  
    \renewcommand{\arraystretch}{1.5}
    \resizebox{\textwidth}{!}{\setlength{\pdfpagewidth}{8.5in} \setlength{\pdfpageheight}{11in}
\begin{tabular}{l *{9}{d{1.3}}} \hline
\multicolumn{1}{c}{} & \multicolumn{3}{c}{OLS} & \multicolumn{3}{c}{1st Stage} & \multicolumn{3}{c}{2SLS} \\ 
\cmidrule(lr){2-4} \cmidrule(lr){5-7} \cmidrule(lr){8-10}
 & \multicolumn{1}{c}{(1)} & \multicolumn{1}{c}{(2)} & \multicolumn{1}{c}{(3)} & \multicolumn{1}{c}{(4)} & \multicolumn{1}{c}{(5)} & \multicolumn{1}{c}{(6)} & \multicolumn{1}{c}{(7)} & \multicolumn{1}{c}{(8)} & \multicolumn{1}{c}{(9)} \\
 & \multicolumn{1}{c}{Log Total Staked} & \multicolumn{1}{c}{Log Total Staked} & \multicolumn{1}{c}{Log Total Staked} & \multicolumn{1}{c}{Log Rewards (USD)} & \multicolumn{1}{c}{Log Rewards (USD)} & \multicolumn{1}{c}{Log Rewards (USD)} & \multicolumn{1}{c}{Log Total Staked} & \multicolumn{1}{c}{Log Total Staked} & \multicolumn{1}{c}{Log Total Staked} \\
 & \multicolumn{1}{c}{(USD)} & \multicolumn{1}{c}{(USD)} & \multicolumn{1}{c}{(USD)} & \multicolumn{1}{c}{} & \multicolumn{1}{c}{1-day Lag} & \multicolumn{1}{c}{7-day Lag} & \multicolumn{1}{c}{(USD)} & \multicolumn{1}{c}{(USD)} & \multicolumn{1}{c}{(USD)} \\ \hline
Log Rewards (USD) & 0.678^{***} &  &  &  &  &  & 1.184^{***} &  &  \\
 & (0.067) &  &  &  &  &  & (0.073) &  &  \\
Log Rewards (USD), 1-day Lag &  & 0.658^{***} &  &  &  &  &  & 1.176^{***} &  \\
 &  & (0.071) &  &  &  &  &  & (0.074) &  \\
Log Rewards (USD), 7-day Lag &  &  & 0.587^{***} &  &  &  &  &  & 1.146^{***} \\
 &  &  & (0.071) &  &  &  &  &  & (0.078) \\
Log Gas Fees (USD) &  &  &  & 0.228^{***} &  &  &  &  &  \\
 &  &  &  & (0.009) &  &  &  &  &  \\
Log Gas Fees (USD), 1-day Lag &  &  &  &  & 0.227^{***} &  &  &  &  \\
 &  &  &  &  & (0.009) &  &  &  &  \\
Log Gas Fees (USD), 7-day Lag &  &  &  &  &  & 0.228^{***} &  &  &  \\
 &  &  &  &  &  & (0.009) &  &  &  \\
Constant & 12.835^{***} & 13.086^{***} & 13.931^{***} & 8.464^{***} & 8.471^{***} & 8.458^{***} & 6.774^{***} & 6.868^{***} & 7.240^{***} \\
 & (0.797) & (0.853) & (0.853) & (0.143) & (0.143) & (0.142) & (0.877) & (0.888) & (0.937) \\
 &  &  &  &  &  &  &  &  &  \\
Observations & \multicolumn{1}{c}{622} & \multicolumn{1}{c}{621} & \multicolumn{1}{c}{615} & \multicolumn{1}{c}{622} & \multicolumn{1}{c}{621} & \multicolumn{1}{c}{615} & \multicolumn{1}{c}{622} & \multicolumn{1}{c}{621} & \multicolumn{1}{c}{615} \\
 R-squared & 0.287 & 0.268 & 0.217 & 0.499 & 0.500 & 0.505 & 0.128 & 0.101 & 0.021 \\ \hline
\multicolumn{10}{c}{ Robust standard errors in parentheses} \\
\multicolumn{10}{c}{ *** p$<$0.01, ** p$<$0.05, * p$<$0.1} \\
\end{tabular}}
\end{table}

\begin{table}[H]
    \centering
    \caption{Total Staking, Daily Data}
    \label{tab:total_daily} 
    \tiny  
    \setlength\tabcolsep{4pt}  
    \renewcommand{\arraystretch}{1.5}
    \resizebox{\textwidth}{!}{\setlength{\pdfpagewidth}{8.5in} \setlength{\pdfpageheight}{11in}
\begin{tabular}{l *{9}{d{1.3}}} \hline
\multicolumn{1}{c}{} & \multicolumn{3}{c}{OLS} & \multicolumn{3}{c}{1st Stage} & \multicolumn{3}{c}{2SLS} \\ 
\cmidrule(lr){2-4} \cmidrule(lr){5-7} \cmidrule(lr){8-10}
 & \multicolumn{1}{c}{(1)} & \multicolumn{1}{c}{(2)} & \multicolumn{1}{c}{(3)} & \multicolumn{1}{c}{(4)} & \multicolumn{1}{c}{(5)} & \multicolumn{1}{c}{(6)} & \multicolumn{1}{c}{(7)} & \multicolumn{1}{c}{(8)} & \multicolumn{1}{c}{(9)} \\
 & \multicolumn{1}{c}{Log Total Staked} & \multicolumn{1}{c}{Log Total Staked} & \multicolumn{1}{c}{Log Total Staked} & \multicolumn{1}{c}{Log Rewards (USD)} & \multicolumn{1}{c}{Log Rewards (USD)} & \multicolumn{1}{c}{Log Rewards (USD)} & \multicolumn{1}{c}{Log Total Staked} & \multicolumn{1}{c}{Log Total Staked} & \multicolumn{1}{c}{Log Total Staked} \\
 & \multicolumn{1}{c}{(USD)} & \multicolumn{1}{c}{(USD)} & \multicolumn{1}{c}{(USD)} & \multicolumn{1}{c}{} & \multicolumn{1}{c}{1-day Lag} & \multicolumn{1}{c}{7-day Lag} & \multicolumn{1}{c}{(USD)} & \multicolumn{1}{c}{(USD)} & \multicolumn{1}{c}{(USD)} \\ \hline
Log Rewards (USD) & 1.279^{***} &  &  &  &  &  & 1.078^{***} &  &  \\
 & (0.018) &  &  &  &  &  & (0.035) &  &  \\
Log Rewards (USD), 1-day Lag &  & 1.276^{***} &  &  &  &  &  & 1.075^{***} &  \\
 &  & (0.019) &  &  &  &  &  & (0.036) &  \\
Log Rewards (USD), 7-day Lag &  &  & 1.269^{***} &  &  &  &  &  & 1.069^{***} \\
 &  &  & (0.022) &  &  &  &  &  & (0.039) \\
Log Gas Fees (USD) &  &  &  & 0.406^{***} &  &  &  &  &  \\
 &  &  &  & (0.019) &  &  &  &  &  \\
Log Gas Fees (USD), 1-day Lag &  &  &  &  & 0.405^{***} &  &  &  &  \\
 &  &  &  &  & (0.019) &  &  &  &  \\
Log Gas Fees (USD), 7-day Lag &  &  &  &  &  & 0.407^{***} &  &  &  \\
 &  &  &  &  &  & (0.019) &  &  &  \\
Constant & 4.618^{***} & 4.668^{***} & 4.804^{***} & 9.212^{***} & 9.226^{***} & 9.190^{***} & 7.739^{***} & 7.786^{***} & 7.900^{***} \\
 & (0.283) & (0.298) & (0.338) & (0.292) & (0.292) & (0.290) & (0.543) & (0.556) & (0.605) \\
 &  &  &  &  &  &  &  &  &  \\
Observations & \multicolumn{1}{c}{622} & \multicolumn{1}{c}{621} & \multicolumn{1}{c}{615} & \multicolumn{1}{c}{622} & \multicolumn{1}{c}{621} & \multicolumn{1}{c}{615} & \multicolumn{1}{c}{622} & \multicolumn{1}{c}{621} & \multicolumn{1}{c}{615} \\
 R-squared & 0.880 & 0.872 & 0.846 & 0.434 & 0.434 & 0.453 & 0.858 & 0.851 & 0.825 \\ \hline
\multicolumn{10}{c}{ Robust standard errors in parentheses} \\
\multicolumn{10}{c}{ *** p$<$0.01, ** p$<$0.05, * p$<$0.1} \\
\end{tabular}}
\end{table}

\begin{table}[H]
    \centering
    \caption{Solo Staking, Weekly Data}
    \label{tab:solo_staking_weekly} 
    \tiny  
    \setlength\tabcolsep{4pt}  
    \renewcommand{\arraystretch}{1.5}
    \resizebox{\textwidth}{!}{\setlength{\pdfpagewidth}{8.5in} \setlength{\pdfpageheight}{11in}
\begin{tabular}{l *{6}{d{1.3}}} \hline
\multicolumn{1}{c}{} & \multicolumn{2}{c}{OLS} & \multicolumn{2}{c}{1st Stage} & \multicolumn{2}{c}{2SLS} \\ 
\cmidrule(lr){2-3} \cmidrule(lr){4-5} \cmidrule(lr){6-7}
 & \multicolumn{1}{c}{(1)} & \multicolumn{1}{c}{(2)} & \multicolumn{1}{c}{(3)} & \multicolumn{1}{c}{(4)} & \multicolumn{1}{c}{(5)} & \multicolumn{1}{c}{(6)} \\
 & \multicolumn{1}{c}{Log Total Staked} & \multicolumn{1}{c}{Log Total Staked} & \multicolumn{1}{c}{Log Rewards (USD)} & \multicolumn{1}{c}{Log Rewards (USD)} & \multicolumn{1}{c}{Log Total Staked} & \multicolumn{1}{c}{Log Total Staked} \\
 & \multicolumn{1}{c}{(USD)} & \multicolumn{1}{c}{(USD)} & \multicolumn{1}{c}{} & \multicolumn{1}{c}{1-week Lag} & \multicolumn{1}{c}{(USD)} & \multicolumn{1}{c}{(USD)} \\ \hline
Log Rewards (USD) & 0.900^{***} &  &  &  & 1.222^{***} &  \\
 & (0.171) &  &  &  & (0.166) &  \\
Log Rewards (USD), 1-week Lag &  & 0.791^{***} &  &  &  & 1.184^{***} \\
 &  & (0.181) &  &  &  & (0.178) \\
Log Gas Fees (USD) &  &  & 0.246^{***} &  &  &  \\
 &  &  & (0.015) &  &  &  \\
Log Gas Fees (USD), 1-week Lag &  &  &  & 0.246^{***} &  &  \\
 &  &  &  & (0.015) &  &  \\
Constant & 10.367^{***} & 11.889^{***} & 10.129^{***} & 10.127^{***} & 5.871^{**} & 6.412^{***} \\
 & (2.376) & (2.520) & (0.230) & (0.229) & (2.313) & (2.479) \\
 &  &  &  &  &  &  \\
Observations & \multicolumn{1}{c}{88} & \multicolumn{1}{c}{87} & \multicolumn{1}{c}{88} & \multicolumn{1}{c}{87} & \multicolumn{1}{c}{88} & \multicolumn{1}{c}{87} \\
 R-squared & 0.355 & 0.281 & 0.751 & 0.755 & 0.309 & 0.212 \\ \hline
\multicolumn{7}{c}{ Robust standard errors in parentheses} \\
\multicolumn{7}{c}{ *** p$<$0.01, ** p$<$0.05, * p$<$0.1} \\
\end{tabular}}
\end{table}

\begin{table}[H]
    \centering
    \caption{Total Staking, Weekly Data}
    \label{tab:total_weekly} 
    \tiny  
    \setlength\tabcolsep{4pt}  
    \renewcommand{\arraystretch}{1.5}
    \resizebox{\textwidth}{!}{\setlength{\pdfpagewidth}{8.5in} \setlength{\pdfpageheight}{11in}
\begin{tabular}{l *{6}{d{1.3}}} \hline
\multicolumn{1}{c}{} & \multicolumn{2}{c}{OLS} & \multicolumn{2}{c}{1st Stage} & \multicolumn{2}{c}{2SLS} \\ 
\cmidrule(lr){2-3} \cmidrule(lr){4-5} \cmidrule(lr){6-7}
 & \multicolumn{1}{c}{(1)} & \multicolumn{1}{c}{(2)} & \multicolumn{1}{c}{(3)} & \multicolumn{1}{c}{(4)} & \multicolumn{1}{c}{(5)} & \multicolumn{1}{c}{(6)} \\
 & \multicolumn{1}{c}{Log Total Staked} & \multicolumn{1}{c}{Log Total Staked} & \multicolumn{1}{c}{Log Rewards (USD)} & \multicolumn{1}{c}{Log Rewards (USD)} & \multicolumn{1}{c}{Log Total Staked} & \multicolumn{1}{c}{Log Total Staked} \\ 
 & \multicolumn{1}{c}{(USD)} & \multicolumn{1}{c}{(USD)} & \multicolumn{1}{c}{} & \multicolumn{1}{c}{1-week Lag} & \multicolumn{1}{c}{(USD)} & \multicolumn{1}{c}{(USD)} \\ \hline
Log Rewards (USD) & 1.338^{***} &  &  &  & 1.085^{***} &  \\
 & (0.028) &  &  &  & (0.081) &  \\
Log Rewards (USD), 1-week Lag &  & 1.327^{***} &  &  &  & 1.078^{***} \\
 &  & (0.039) &  &  &  & (0.090) \\
Log Gas Fees (USD) &  &  & 0.452^{***} &  &  &  \\
 &  &  & (0.048) &  &  &  \\
Log Gas Fees (USD), 1-week Lag &  &  &  & 0.453^{***} &  &  \\
 &  &  &  & (0.048) &  &  \\
Constant & 3.041^{***} & 3.257^{***} & 10.439^{***} & 10.428^{***} & 7.468^{***} & 7.610^{***} \\
 & (0.482) & (0.675) & (0.753) & (0.753) & (1.404) & (1.563) \\
 &  &  &  &  &  &  \\
Observations & \multicolumn{1}{c}{88} & \multicolumn{1}{c}{87} & \multicolumn{1}{c}{88} & \multicolumn{1}{c}{87} & \multicolumn{1}{c}{88} & \multicolumn{1}{c}{87} \\
 R-squared & 0.913 & 0.883 & 0.506 & 0.528 & 0.880 & 0.852 \\ \hline
\multicolumn{7}{c}{ Robust standard errors in parentheses} \\
\multicolumn{7}{c}{ *** p$<$0.01, ** p$<$0.05, * p$<$0.1} \\
\end{tabular}
}
\end{table}

\newpage 

\section{Discussion}\label{discussion}

There are several limitations to the theoretical model developed in this paper. First, it is important to note that the model does not consider entry or exits of staking agents, i.e., we assume a fixed number of agents per ETH holder type. Further, the model does not allow stakers to switch between different staking methods. Based on these assumptions, the model framework may be able to model the short-run equilibrium dynamics of the Ethereum staking market, but may fall short when considering the long-run perspective. We argue that the results presented above primarily reflect a short-term perspective on how the staking market might respond to changes in the issuance schedule. In the long run, the relative profitability of different staking solutions is likely to influence market shares, as new entrants or existing market participants opt for profitable staking methods and existing stakers exit unprofitable ones.

\subsection{Short-term Perspective}

As argued above, the empirical estimates can be argued to be broadly consistent with the outcomes of the calibrated game-theoretic model, where solo stakers are modeled to experience disutility from variability in MEV revenues and where a portion of individuals staking through centralized providers are assumed to be inattentive. Under these assumptions, the staking supply of solo stakers is shown to adjust more strongly to changes in the issuance schedule compared to other types of stakers. Consequently, under the proposed issuance schedule, the market share of solo staking would be expected to decline, while the market share of centralized exchanges is shown to increase. 

This finding has important implications for the broader discussion around modifying the Ethereum issuance schedule. On the one hand, it suggests that a reduction in issuance could help limit the market share of liquid staking providers and the resulting dominance of LSTs within the Ethereum ecosystem. On the other hand, a reduction in the issuance schedule may come at the cost of an increased market share for centralized providers. In combination with a declining market share of solo stakers, this could contribute to a greater degree of centralization in the Ethereum validator set.

Finally, the results underscore that even with cost structures that would otherwise make solo stakers inelastic to changes in consensus yields, the competitive environment in the broader staking market can still make solo stakers vulnerable to being crowded out of the staking market when consensus yields fall. This is because even in a segregated staking market where existing solo stakers do not switch to other staking solutions at all, other market participants may be relatively insensitive to changes in consensus issuance and thus put pressure on solo stakers to reduce their staking supply when consensus yields fall. 

This result also has important implications for the broader economic policy mix of the Ethereum protocol. It suggests that policies that help improve the competitive standing of solo stakers relative to other staking methods, for example the implementation of MEV burn, could help mitigate the otherwise negative effects of an issuance reduction on the market share of solo stakers.

\subsection{Long-term Perspective}

Across all model specifications evaluated above, a change in the issuance schedule is associated with a disproportionately negative impact on the expected profitability of solo staking relative to other staking methods. Over time, this short-run decline in profitability may increase incentives for solo stakers to either exit the market or transition to alternative staking methods. On the other hand, the profitability of \textit{Techies} staking via dSSPs that offer additional DeFi yields is least affected by the decline in issuance, particularly also in comparison to staking via centralized exchanges. In the long run, individual LSTs (or LRTs) may therefore continue to gain market share. The long-term effects on the centralization of the Ethereum validator set are thus ambiguous. Based on the results above, both the most decentralized staking option (solo staking) and the most centralized option (staking via CEXs) become less attractive in the long run.

\begin{figure}[H]
    \caption{Cumulative long-run effects of profit reinvestment}
    \centering
    \label{fig:cumulative_profits}
    \begin{subfigure}[t]{0.45\textwidth}  
        \centering
        \includegraphics[width=\linewidth]{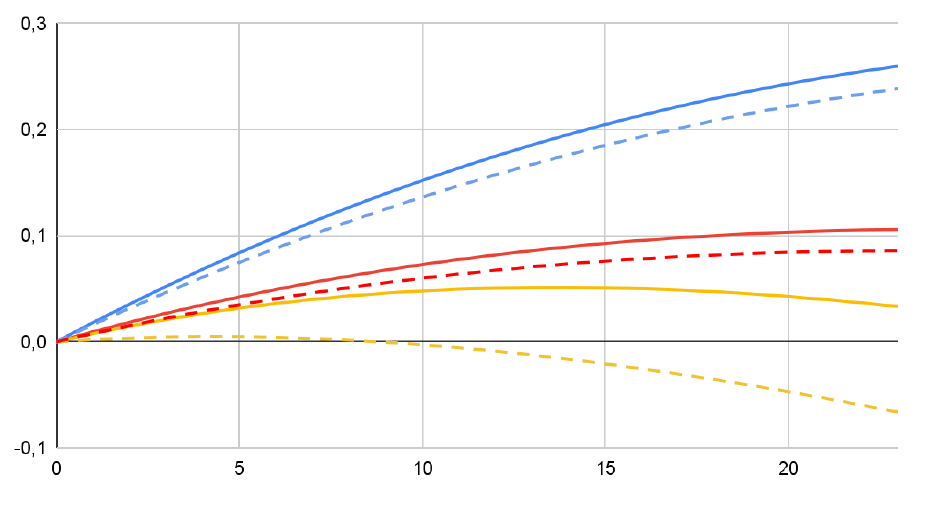}
        \caption{Staking level}
        \label{fig:sub-first}
    \end{subfigure}%
    \hfill
    \begin{subfigure}[t]{0.45\textwidth}  
        \centering
        \includegraphics[width=\linewidth]{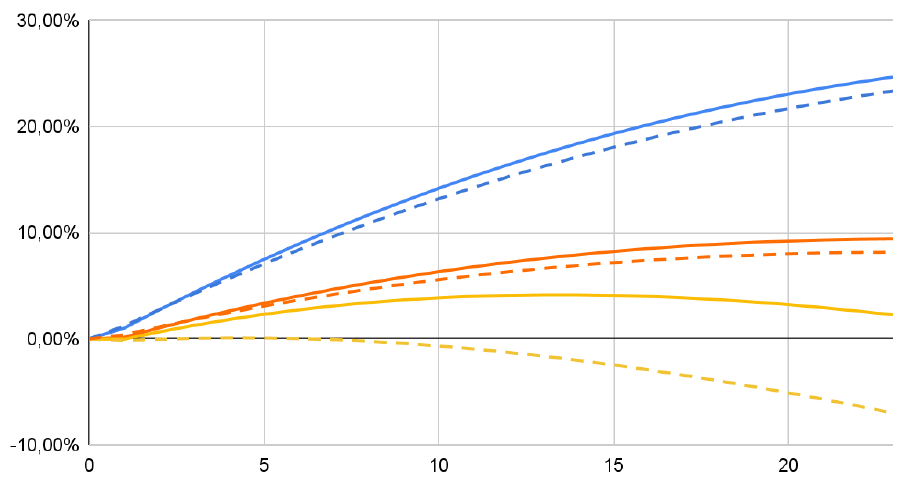}
        \caption{Staking ratio}
        \label{fig:sub-second}
    \end{subfigure}%
    \caption*{\footnotesize \textit{Note}: Full line represents current issuance yield; dashed line represents adjusted issuance yield; Blue line represents Techies; Red represents Retailers; Yellow represents solo stakers.}
\end{figure}

However, exits and market participants switching between staking methods as a result of diminished profitability are not the only factors that could lead to diminished solo staking and an increasing dominance of liquid staking solutions in the market. This is because, over time, the divergence in profits between the different staking methods leads to a redistributive effect between the different staking methods. If stakers are able to reinvest their profits from staking, and staking profits are comparatively low for solo staking, their market share will continue to diminish. To illustrate this effect, we use the Nash equilibria of the baseline model outlined in Table \ref{tab:staking_summary_baseline} as a starting point and chart the cumulative changes in staking levels and staking ratios over time, assuming stakers reinvest their profits (revenues minus costs) and do not reevaluate their optimal level of staking supply over time. The results are shown in Figure \ref{fig:cumulative_profits}. Figure \ref{fig:sub-first} depicts the cumulative percentage change in the total staking level relative to the starting point. Similarly, Figure \ref{fig:sub-second} depicts the cumulative change in the share of total ETH supply being staked. As such, Figure \ref{fig:sub-second} also takes into account the effects of dilution. 

This thought experiment illustrates the potential redistributive effects arising from differences in profitability, ultimately leading to the gradual displacement of solo stakers from the market over the long term. Moreover, the calculations show that a reduction in the consensus issuance schedule noticeably accelerates this process, as the profitability of solo stakers is disproportionally affected. How strong this effect is in reality depends to a large degree on the extent to which stakers actually reinvest their staking profits. Anecdotal evidence suggests that ETH holders who stake through intermediaries are significantly better positioned to reinvest profits compared to small solo stakers. Given that the protocol sets a maximum effective balance for validators, smaller solo stakers encounter practical difficulties in expanding their revenue base through the reinvestment of staking rewards. In contrast, ETH holders utilizing intermediaries can more effectively compound their returns due to the ability to pool staking profits collectively.

\section{Conclusion}\label{conclusion}

Through empirical analysis and a game-theoretic model, we show that solo stakers are more responsive to changes in consensus issuance yields than ETH holders staking through centralized exchanges or liquid staking providers. As issuance declines, solo staking is expected to lose market share, while centralized exchanges gain dominance.

This heightened sensitivity of solo stakers persists despite their cost structures and is instead primarily driven by the competitive dynamics of the staking market. Other staking methods benefit from superior MEV access and DeFi yields, making them less responsive to issuance changes and creating a crowding-out effect on solo stakers. These findings underscore the importance of mechanisms such as MEV burn in mitigating the negative impact of issuance reductions on solo stakers.

Additionally, reduced issuance disproportionately lowers solo staking profitability and hence the incentives for solo staking which may drive exits or shifts toward alternative staking solutions like dSSPs with DeFi yields over time.

The analysis presented in this paper has limitations. In particular, ike the assumptions used in calibrating the model. While the assumptions used in calibrating the theoretical model align with existing literature, they remain open to debate. Future research should prioritize refining the estimation of cost structures across different staking methods, which could enhance the accuracy and robustness of the model framework. Another important set of assumptions is that stakers are homogeneous within each type and that the ETH price remains constant over time (it is not affected by inflation and/or by the liquidity level). Relaxing these assumptions represents a promising direction for future research.

We expect that the model developed here serves as a foundation for further research and extensions, providing a flexible framework for analyzing staking dynamics under various protocol adjustments. For tractability and compatibility with the broader model, we introduced several simplifying assumptions in the model extension with endogenous dSSP fees. However, we believe that our model could be expanded to facilitate a deeper exploration of the competitive dynamics between staking intermediaries.


\bibliographystyle{apalike}            
\bibliography{references}             

\appendix  

\newpage

\section{Derivation of Theoretical Model}\label{sec:derivation}

\textbf{Baseline model:}

We solve for the equilibrium staking supply in the baseline model under the current issuance schedule ( the same approach is used to solve the extended models). In this case, the yield curve is:
        \begin{equation*}
            y_i(D_T) = \frac{cF}{\sqrt{D_T}}=\frac{2.6\cdot 64}{\sqrt{D_T}}
        \end{equation*}

\noindent Each staker decides how much to stake taking into account other stakers' behavior.  Thus, we solve equations (\ref{eq:solo_staker})-(\ref{eq:retailer}), getting
\begin{align*}
   \frac{\partial y_i(D)}{\partial \hat{d}_{ss}}
& \cdot \hat{d}_{ss}+y_i(D)+y_v\cdot \frac{\partial P_{ss}(D) }{\partial \hat{d}_{ss}}-c_{ss}\cdot \frac{\partial \hat{d}_{ss}^{\alpha_{ss}}}{\partial \hat{d}_{ss}}=0,\\
 (1-f_d)&\cdot\left[\frac{\partial y_i(D)}{\partial \hat{d}_{t}}
\cdot \hat{d}_{t}+y_i(D)+y_v\cdot \frac{\partial P_{t}(D) }{\partial \hat{d}_{t}}\right]+y_d-c_{t}\cdot \frac{\partial \hat{d}_{t}^{\alpha_{t}}}{\partial \hat{d}_{t}}=0,\\
 (1-f_c)&\cdot\left[\frac{\partial y_i(D)}{\partial \hat{d}_{r}}
\cdot \hat{d}_{r}+y_i(D)+y_v\cdot \frac{\partial P_{r}(D) }{\partial \hat{d}_{r}}\right]-c_{r}\cdot \frac{\partial \hat{d}_{r}^{\alpha_{r}}}{\partial \hat{d}_{r}}=0,
\end{align*} 
where
\begin{align*}
     y_i(D)&=\frac{2.6\cdot 64}{\sqrt{\hat{d}_{\theta}+(N_{\theta}-1)\cdot d_{\theta}+\sum_{-\theta}D_{-\theta}}},\\   
   \frac{\partial y_i(D)}{\partial \hat{d}_{\theta}}\cdot \hat{d}_{\theta}&=-\frac{1}{2}\cdot \frac{2.6\cdot 64}{(\hat{d}_{\theta}+(N_{\theta}-1)\cdot d_{\theta}+\sum_{-\theta}D_{-\theta})^{(3/2)}} \cdot \hat{d}_{\theta},\\
     y_v\cdot \frac{\partial P_{\theta}(D) }{\partial \hat{d}_{\theta}}&=y_v\cdot\frac{(N_{\theta}-1)\cdot d_{\theta}+\sum_{-\theta}D_{-\theta}}{(\hat{d}_{\theta}+(N_{\theta}-1)\cdot d_{\theta}+\sum_{-\theta}D_{-\theta})^{2}},\\
     c_{\theta}\cdot \frac{\partial \hat{d}_{\theta}^{\alpha_{\theta}}}{\partial \hat{d}_{\theta}}&=    c_{\theta}\cdot\alpha_{\theta}\cdot \hat{d}_{\theta}^{\alpha_{\theta}-1}.
\end{align*}
The first three terms converge to zero for any  $\hat{d}_{\theta}<<(N_{\theta}-1)\cdot d_{\theta}+\sum_{-\theta}D_{-\theta}$ when $(N_{\theta}-1)\cdot d_{\theta}+\sum_{-\theta}D_{-\theta}$ growth.  However,  $\frac{\partial y_i(D)}{\partial \hat{d}_{\theta}}\cdot \hat{d}_{\theta}$  converges faster than $y_i(D)$ and $y_v\cdot \frac{\partial P_{\theta}(D) }{\partial \hat{d}_{\theta}}$. Additionally,  
\begin{equation*}
    \frac{(N_{\theta}-1)\cdot d_{\theta}+\sum_{-\theta}D_{-\theta}}{(\hat{d}_{\theta}+(N_{\theta}-1)\cdot d_{\theta}+\sum_{-\theta}D_{-\theta})^{2}}\approx  \frac{1}{(\hat{d}_{\theta}+(N_{\theta}-1)\cdot d_{\theta}+\sum_{-\theta}D_{-\theta})}=\frac{P_{\theta}(D)}{\hat{d}_{\theta}}.
\end{equation*}

\newpage
\noindent These points are exemplified in the following figures:

 \begin{figure}[H]
      \caption{Illustration of Simplifying Assumptions}
    \centering
    \includegraphics[width=0.6\linewidth]{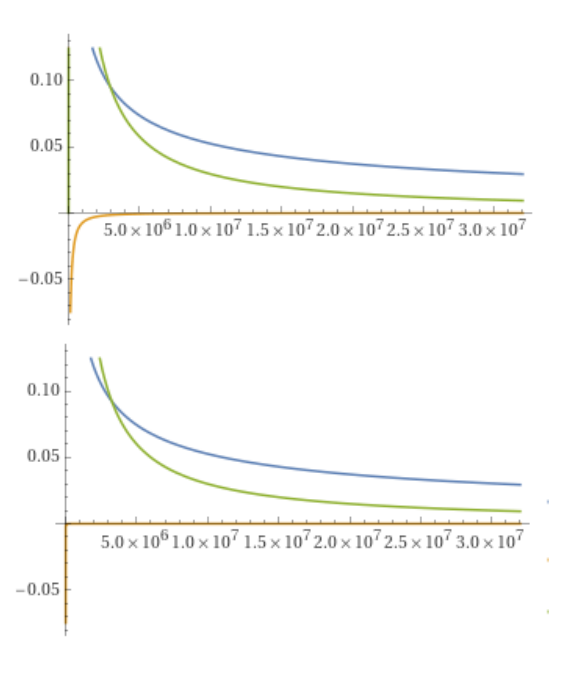}
    \caption*{\footnotesize \textit{Note:} x-Axis: Staking levels. y-Axis: $y_i(D)$ (blue ), $y_v\cdot P_{\theta}'(D)$ (green), and $y_i'(D)\cdot \hat{d}_{\theta}$ (orange). The figure above has $\hat{d}_{\theta}=100K$, while below $\hat{d}_{\theta}=32$. }
    \end{figure}

\begin{figure}[H]
      \caption{Illustration of Simplifying Assumptions}
    \centering\includegraphics[width=0.6\linewidth]{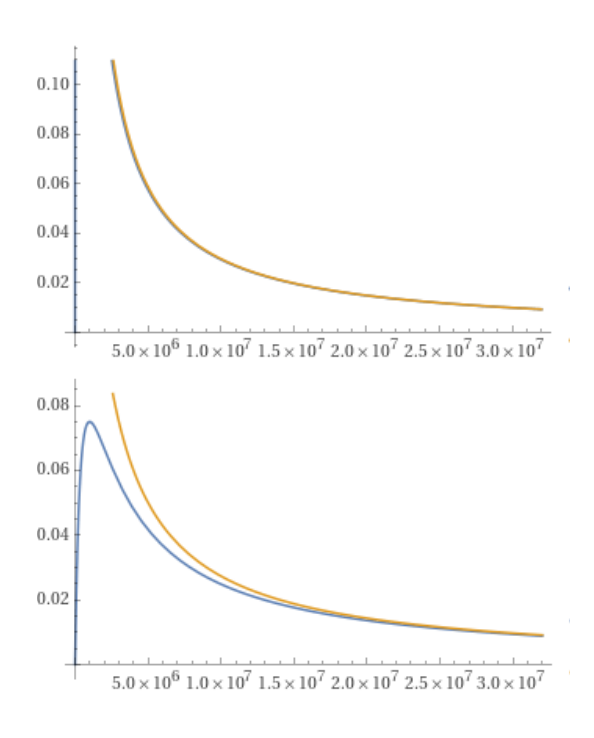}
    \caption*{\footnotesize \textit{Note:} x-Axis: Staking levels. y-Axis: $y_v\cdot P_{\theta}'(D)$ (blue) and $y_v\cdot P_{\theta}(D)/\hat{d}_{\theta}$ (orange). The figure above uses $\hat{d}_{\theta}=100K$, and below $\hat{d}_{\theta}=1'000K$. }
    \end{figure}

\noindent The previous observation allows us to simplify the FOCs and using $\hat{d}_{\theta}=d_{\theta}$ for each type in equilibrium, then the system of equations is, which can be solved numerically:
 \begin{align*}
  &  y_i(D)+y_v\cdot \frac{P_{ss}(D)}{d_{ss}}=c_{ss}\cdot \frac{\partial d_{ss}^{\alpha_{ss}}}{\partial d_{ss}},\quad \text{for each ss}\\
 & (1-f_d)\cdot \left[y_i(D)+y_v\cdot \frac{P_{t}(D)}{d_{t}}\right]+y_d=c_{t}\cdot \frac{\partial d_{t}^{\alpha_{t}}}{\partial d_{t}},\quad \text{for each t}\\
& (1-f_c)\cdot \left[y_i(D)+y_v\cdot \frac{P_{r}(D)}{d_{r}}\right]=c_{r}\cdot \frac{\partial d_{r}^{\alpha_{r}}}{\partial d_{r}}, \quad \text{for each r}.
\end{align*}

\textbf{Solos Stakers Expected MEV Revenues:}

To model the variance over MEV rewards for moderate risk-averse solo-stakers, we can make

\begin{equation}\label{equation_SS_problem_with_variance}
         \max_{\hat{d}_{ss}} \quad y_i(D) \cdot \hat{d}_{ss}+[y_v\cdot P_{ss}(D)-\frac{1}{2 } \cdot y_v^2\cdot P_{ss}(D)\cdot (1-P_{ss}(D))]-(C_{ss}+c_{ss}\cdot \hat{d}_{ss}^{\alpha_{ss}}),
\end{equation}
where we are using
\begin{align*}
    E[Y_{MEV}]-Var[Y_{MEV}]&=y_v\cdot P_{ss}(D)-\frac{1}{2}\cdot y_v^2\cdot P_{ss}(D)\cdot (1-P_{ss}(D)).
\end{align*}
Notice that, since $P_{ss}(D)=\frac{d_{ss}}{D}\longrightarrow 0$ for small $d_{ss}$, then $E[Y_{MEV}]-Var[Y_{MEV}]\approx 0$. It follows that when we account for the higher variance in MEV rewards faced by solo stakers, a moderately risk-averse and small Solo Staker is assumed to exclude these revenues when deciding how much to stake, solving
\begin{equation}\label{equation_SS_problem_with_variance}
         \max_{\hat{d}_{ss}}\quad y_i(D) \cdot \hat{d}_{ss}-(C_{ss}+c_{ss}\cdot \hat{d}_{ss}^{\alpha_{ss}}).
\end{equation}
Rather, we consider in the following a less risk-averse solo staker, with

\begin{equation*}
    E[Y_{MEV}]-Var[Y_{MEV}]= y_v\cdot P_{ss}(D)- y_v^2\cdot P_{ss}(D)^2,
\end{equation*}
and for the first-order condition, we take the following approximation:
\begin{equation*}
    \frac{\partial [y_v\cdot P_{ss}(D)-y_v^2\cdot P_{ss}(D)^2]}{\partial d_{ss}}\approx y_v \cdot \frac{P_{ss}(D)}{d_{ss}}\cdot (1-2\cdot y_v \cdot P_{ss}(D))
\end{equation*}
\end{document}